\documentclass[twocolumn, trackchanges]{aastex701}
\usepackage{graphicx}
\usepackage{bm}
\usepackage{amssymb}
\usepackage{amsmath}
\usepackage{units}
\usepackage{array}

\usepackage[T1]{fontenc}
\usepackage{aecompl}
\usepackage{calligra}
\usepackage{soul}
\DeclareMathAlphabet{\mathcalligra}{T1}{calligra}{m}{n}
\DeclareFontShape{T1}{calligra}{m}{n}{<->s*[2.2]callig15}{}

\newcommand{\Mdot}[0]{\dot M}

\newcommand{\Mach}[0]{\mathcal{M}}

\newcommand{\edd}[0]{{\rm Edd}}
\newcommand{\eff}[0]{{\rm eff}}
\newcommand{\feff}[0]{f_{\rm eff}}
\newcommand{\crit}[0]{{\rm crit}}
\newcommand{\Msun}[0]{{M_\odot}}

\newcommand{\sink}[0]{{\rm sink}}

\newcommand{\nuem}[0]{{\nu_{\rm em}}}
\newcommand{\nuobs}[0]{{\nu_{\rm obs}}}
\newcommand{\nusen}[0]{{\nu_{\rm sen}}}
\def\eg{{e.g., }}

\usepackage[dvipsnames]{xcolor}

\usepackage{color}

\usepackage{booktabs}

\submitjournal{\apj}
\received{June 4, 2026}
\revised{July 23, 2026}
\accepted{July 25, 2026}

\shorttitle{Big binary -- cold disk}
\shortauthors{Tiede, O'Neill \& D'Orazio}
\graphicspath{{./}{Figures/}}

\begin{document}

\title{Dynamics and detectability of long-lived non-accretion phases for massive black hole binaries in cold, thermally regulating disks}

\correspondingauthor{Christopher Tiede}
\email{christopher.tiede@nbi.ku.dk}
\author[0000-0002-3820-2404]{Christopher Tiede}
\affiliation{Niels Bohr International Academy, Niels Bohr Institute, Blegdamsvej 17, 2100 Copenhagen, Denmark}
\email{christopher.tiede@nbi.ku.dk}
\author[0000-0002-1382-3802]{David O'Neill}
\affiliation{Institute of Science and Technology Austria (ISTA), Am Campus 1, Klosterneuburg 3400, Austria}
\affiliation{Niels Bohr International Academy, Niels Bohr Institute, Blegdamsvej 17, 2100 Copenhagen, Denmark}
\email{David.O’Neill@ist.ac.at}
\author[0000-0002-1271-6247]{Daniel J. D'Orazio}
\affiliation{Space Telescope Science Institute, 3700 San Martin Drive, Baltimore, MD 21218, USA}
\affiliation{Department of Physics and Astronomy, Johns Hopkins University, 3400 North Charles Street, Baltimore, Maryland 21218, USA}
\affiliation{Niels Bohr International Academy, Niels Bohr Institute, Blegdamsvej 17, 2100 Copenhagen, Denmark}
\email{dorazio@stsci.edu}
%

%
% =============================================================================
\begin{abstract}
    % \noindent
    We investigate whether the non-accreting phases found in thin, locally isothermal circumbinary disks survive when the disk thermodynamics are evolved self-consistently. 
    We present grid-based hydrodynamics simulations of circumbinary accretion with an energy equation that includes viscous and hydrodynamic heating coupled to radiative blackbody cooling in the high-Mach number regime. 
    We find that, although gas accumulates and heats at the far edge of the circumbinary cavity, the regions that launch accretion streams remain comparatively cold, leading to potentially long-lived suppression of the binary accretion rate as the large-scale feeding rate is reduced towards the Eddington limit. 
    This runaway non-accretion problem, however, is weakened relative to locally isothermal solutions. 
    Despite their low accretion rates, binaries interacting with disks in a non-accreting phase can remain sufficiently luminous and variable at optical and near-infrared frequencies to be detectable in upcoming wide-field surveys like LSST and the Roman Space Telescope. 
    Because of the effective truncation of the surrounding disk, though, such systems are comparatively faint in high energy, photo-ionizing emission, and may therefore appear as intrinsically X-ray-weak AGN with weak or absent emission line features.
    We additionally suggest an update to grid-based sink prescriptions for approximating mass loss across an unresolved horizon when including an energy conservation equation.
\end{abstract}

%
% ------------------------------------------------------
%

%
% ======================================================================
\section{Introduction} \label{S:intro}
\setcounter{footnote}{0}

The mergers of galaxies hosting super-massive black holes are expected to create super-massive black hole binaries \citep[SMBHBs;][]{Begel:Blan:Rees:1980, Volonteri+2003}. Just as single super-massive black holes interact with extant reservoirs of galactic gas as standard active galactic nuclei \citep[AGN;][]{Gaskell:1985, BarnesHernquist1996}, many SMBHBs are anticipated to undergo phases of gas accretion.
The community has long sought features in AGN emission that signal the presence of an unresolved inner binary, with the most typical being periodic variability associated to a binary's orbital motion \citep[see e.g.,][for reviews]{Komossa:review:2006, bogdanovicLRR+2022, DOrazioCharisi:Review:2023}.
Such features would be critical for the identification of SMBHBs in electromagnetic (EM) surveys \citep[\eg][]{Graham+2015, Charisi:2016, Liu+2019, ChenXin:2020}, enabling both characterization of the full SMBHB population and possible multi-messenger detections in gravitational waves from pulsar timing experiments or future space-based interferometers like LISA.

It is typically assumed that SMBHBs accrete at rates similar to standard AGN, because, while the rotating binary potential alters the structure of the inner accretion flow \citep{al94, Cuadra2009, Shi+2012}, (magneto-) hydrodynamics simulations have demonstrated that gas still falls onto the component black holes via tidal streams \citep[e.g.,][]{MacFadyen2008, Shi2015}.
This transfer of mass onto each component BH enables mass flow in the disk to equilibrate with that onto the binary \citep{Dorazio2013, MML17, MML19}.
Simulations that self-consistently evolve the heating (shocks, compression, and viscosity) and cooling (thermal blackbody radiation) of the solution through an energy conservation equation, have also generally reinforced the facsimile between the average EM appearance of accreting SMBHBs and single SMBHs \citep{Farris15, dAscoliNoble:2018, Westernacher-Schneider:2022, Cocchiararo:2024}.
However, studies focusing on the final gravitational wave driven inspiral phase have noted possible macroscopic signatures like secular luminosity evolution, late-time X-ray turn-off, and post-merger jet launching \citep{Noble+2012, Krauth:2023, Avara:2024, ONeillTiede:2025, EnnoggiKrolik+2025, Zrake:CLIs:2025}

The majority of the above mentioned work, though, was performed for disks with relatively large amounts of pressure support, parametrized through the disk aspect-ratio, $h/r \sim 0.1$.
While this value, in standard disk modeling \citep{SS1973}, is applicable to steady-state disks around stellar mass objects, the typical aspect-ratio around SMBHs would be nearly two orders of magnitude thinner (see \eg Appendix A in \citealt{Valli:2024}; though for qualitatively different configurations see \eg \citealt{Hopkins:MAD-Quasars:2024}).
Equivalently, in the same disk modeling, if one fixes the central mass $M$, viscosity parameter $\alpha$, and aspect-ratio, the functional Eddington fraction of the system scales like $ \propto (h/r)^5$,
such that over-estimating $h/r$ by a factor of $10-100$ corresponds to dramatic changes in the implied astrophysical configuration.
\hspace{-3pt}\footnote{The quoted scaling is for a gas-pressure dominated disk with dominant opacity from electron scattering, but in the free-free opacity regime, the scaling becomes $\propto (h/r)^{20/3}$.}

Multiple studies have now demonstrated that the reduced pressure support in systems with $h/r < 0.1$ meaningfully alters the circumbinary accretion dynamics \citep{Ragusa+2016, Tiede2020, Heath+Nixon2020, Penzlin2022, Franchini2022, Dittmann2022, Dittmann:2026, ClyburnZrake:2026}, and that in the canonical limit of $h/r \sim 10^{-2} - 10^{-3}$, accretion onto the binary may functionally cease \citep{Tiede:2025, Betancourt:2026}. 
This truncation could cause SMBHBs to appear electromagnetically distinct from typical AGN, but
it has been suggested that the extra heat from shocks and tidal effects in the complex inner accretion flow may warm circumbinary disks back to the fiducial thickness, $h/r \approx 0.1$ \citep{Cocchiararo:2024}.
However, no study has yet systematically explored this classically thin limit with the inclusion of an energy equation.

In this work, we report high-resolution hydrodynamics simulations of circumbinary accretion disks with $h/r < 0.1$ that consistently include hydrodynamic heating and radiative blackbody cooling.
We demonstrate that long-lived non-accretion phases, observed in isothermal simulations, persist with this more complete treatment, because the disk remains cold at its inner-most edge.
We document our numerical approach, system setup, and single black hole test simulations in Section~\ref{S:methods}.
We evaluate the dynamics of binary accretion in such classically thin disks in Section~\ref{S:results}, and we report the associated broadband thermal emission properties in Section~\ref{S:emission}.
We discuss implications for super-massive binary populations, ramifications for electromagnetic searches, numerical limitations, future directions, and our main conclusions in Sections~\ref{S:discussion}~and~\ref{S:conclusions}.

%
% ======================================================================
\section{Numerical Methods} \label{S:methods}
%

%
% ------------------------------------------------------
\subsection{System setup} \label{s:system-setup}

We follow the general setup from \citet{Tiede:2025} in solving the vertically averaged, 2D compressible Navier-Stokes Equations for a thin gas disk around a binary of total mass $M$, semi-major axis $a$, and orbital frequency $\Omega_b$, with the finite volume code \texttt{Sailfish} \citep{Sailfish:2024}.
We consider a disk of surface density $\Sigma$ in vertical hydrostatic equilibrium with characteristic Mach number, $\Mach \approx (h/r)^{-1}$, where $h$ is the disk scale-height at radius $r$.
In addition to conserving mass and momentum in our solutions, following \citet{Westernacher-Schneider:2022}, we evolve an energy equation that self-consistently tracks the gas heating and cooling.
We adopt an ideal gas equation of state relating the specific internal energy $\epsilon$ to the vertically integrated pressure, $P = \epsilon \Sigma (\gamma - 1)$ with adiabatic index $\gamma = 5/3$.
Heat injection from dissipation is included through a source term, $\mathbf{\nabla} \cdot \left( \mathbf{v} \cdot \mathbf{\tau}\right)$ for mid-plane fluid velocity $v_i$ and viscous stress tensor $\tau_{ij} = \nu \Sigma \big( \partial_i v_j + \partial_j v_i - (2/3) \delta_{ij}\partial_k v_k \big)$.
We compute the kinematic viscosity $\nu = \alpha c_s h$ through the dimensionless number $\alpha$ and the local adiabatic sound speed $c_s = \sqrt{\gamma P / \Sigma}$.
We assume that our solution is everywhere optically thick,\footnote{We verify that this assumption is generally true, although we examine it in more detail in Section~\ref{S:emission}} 
such that energy losses occur through radiative blackbody cooling at a rate
\begin{align}
    \Lambda = \frac{8}{3} \frac{\sigma T^4}{\kappa \Sigma} \ ,
 \label{eq:lambda-cool}
\end{align}
where $T = (m_p / k_b) P / \Sigma$ is the mid-plane temperature (assuming a gas density dominated by hydrogen with mass $m_p$), 
$\sigma$ and $k_b$ denote the usual Stefan-Boltzmann and Boltzmann constants, respectively, and we assume that the gas opacity is dominated by electron scattering, $\kappa = 0.4\ \unit{cm}^2\ \unit{g}^{-1}$.
The cooling term is applied in a semi-implicit manner as described in \citet{RyanMacFadyen2017}.

We consider an accretion flow that circularizes and flattens at scales much larger than $a$. By the time the inner-disk viscously spreads down to scales comparable to $a$, we assume its outer edge is sufficiently far so to approximate the disk as infinite.
Thus, the outer boundary mimics an infinite accretion disk with a steady inflow of material, $\Mdot_\infty$, via a buffer term that smoothly damps the outer solution to the initial configuration \citep[see][for details]{Westernacher-Schneider:2022}.
The inner boundaries are characterized by sink terms that approximate the loss of mass, momentum, and energy across the (unresolved) horizon of each black hole.
We employ torque-free (or spinless) sinks \citep{Dempsey2020, Dittmann+Ryan2021} of characteristic radius $r_{\rm sink} = 0.05a$, but we implement a modified energy sink term---compared to that employed in previous studies evolving an energy equation \citep[e.g,][]{WangBaiLai:2023, Westernacher-Schneider:2023}---that accounts for the injected azimuthal kinetic energy typically associated to the application of the torque-free condition.
We refer to these as \emph{adiabatic torque-free sinks}, and we provide a derivation and demonstration in Appendix~\ref{app:energy-sink}.

%
% ------------------------------------------------------
\subsection{Disk model} \label{S:disk-model}

The initial disk setup is that of a circular, Keplerian disk corrected for radial pressure gradients, and inward radial drift velocity associated to $\dot M_\infty$.
The surface density and pressure are given by a gas-pressure and electron-scattering-opacity dominated $\alpha$-disk with no net angular momentum current \citep{HKM09}.
Typically, one selects the central mass $M$, the viscosity parameter $\alpha$, and the feeding rate $\Mdot_\infty$, such that
\begin{align}
 \label{eq:sigma_disk}
    \Sigma(r\,;\, \alpha, M, \Mdot_\infty) &= \frac{2 }{(9 \pi)^{3/5} } \bigg( \frac{m_p}{\gamma k_b} \bigg)^{4/5} \bigg( \frac{\sigma}{\kappa} \bigg)^{1/5} \\ 
                &\quad \times\ \alpha^{-4/5} (GM)^{1/5} \Mdot_\infty^{3/5}\, r^{-3/5} \nonumber \\
    P(r\, ;\, \alpha, M, \Mdot_\infty) &= \frac{1}{3\pi\gamma} \alpha^{-1} (GM)^{1/2} \Mdot_\infty\, r^{-3/2} \ ,
 \label{eq:pressure_disk}
\end{align}
and the Mach number profile of the flow can be expressed
\begin{align}
    \Mach^2(r) \equiv -\frac{\phi_g}{c_s^2} = \frac{GM\Sigma(r)}{r\gamma P(r)} \ 
\end{align}
for gravitational potential $\phi_g$ corresponding, here, to a single point mass.
We set $\alpha = 0.1$, but in place of selecting $\Mdot_\infty$, we instead choose $\Mach_a \equiv \Mach(a)$ in the single central potential, such that the effective accretion rate at large scales becomes
\begin{equation}
 \begin{aligned}
    \Mdot_\infty \!=\! 4\pi \sqrt{\frac{2}{3}}  \!  \bigg( \frac{m_p^4 \sigma}{ k_b^4 \gamma \kappa} \bigg)^{\! 1/2} \!\!  \alpha^{1/2} (GM)^{7/4} a^{-1/4} \Mach_a^{-5}.
 \end{aligned}
 \label{eq:mdot}
\end{equation}
Our initial surface density and pressure profiles are then
\begin{align}
    \Sigma(r) = \Sigma_a \bigg( \frac{r}{a} \bigg)^{-3/5} \quad \& \quad
    P(r) = P_a \bigg( \frac{r}{a} \bigg)^{-3/2}
\end{align}
with $\Sigma_a \equiv \Sigma(a;\, \alpha, M, \Mach_a)$ and $P_a \equiv P(a;\, \alpha, M, \Mach_a)$.

%
% ------------------------------------------------------
\subsection{Single point-mass solutions} \label{s:single-mass-tests}
\begin{figure}[t!]
    \centering
    \includegraphics{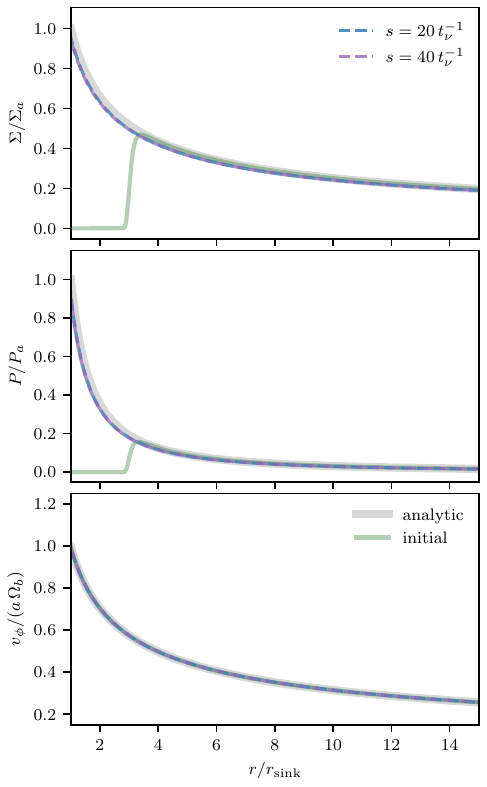}
    \caption{
    Surface density (top), vertically averaged pressure (middle), and orbital velocity (bottom) for a disk around a single point mass for two values of the mass removal rate.
    Grey-bands indicate the analytic $\alpha$-disk solutions, and green lines, the initial profiles.
    }
    \label{fig:single-profiles}
\end{figure}
\begin{figure*}[t!]
    \centering
    \vspace{-5pt}
    \includegraphics{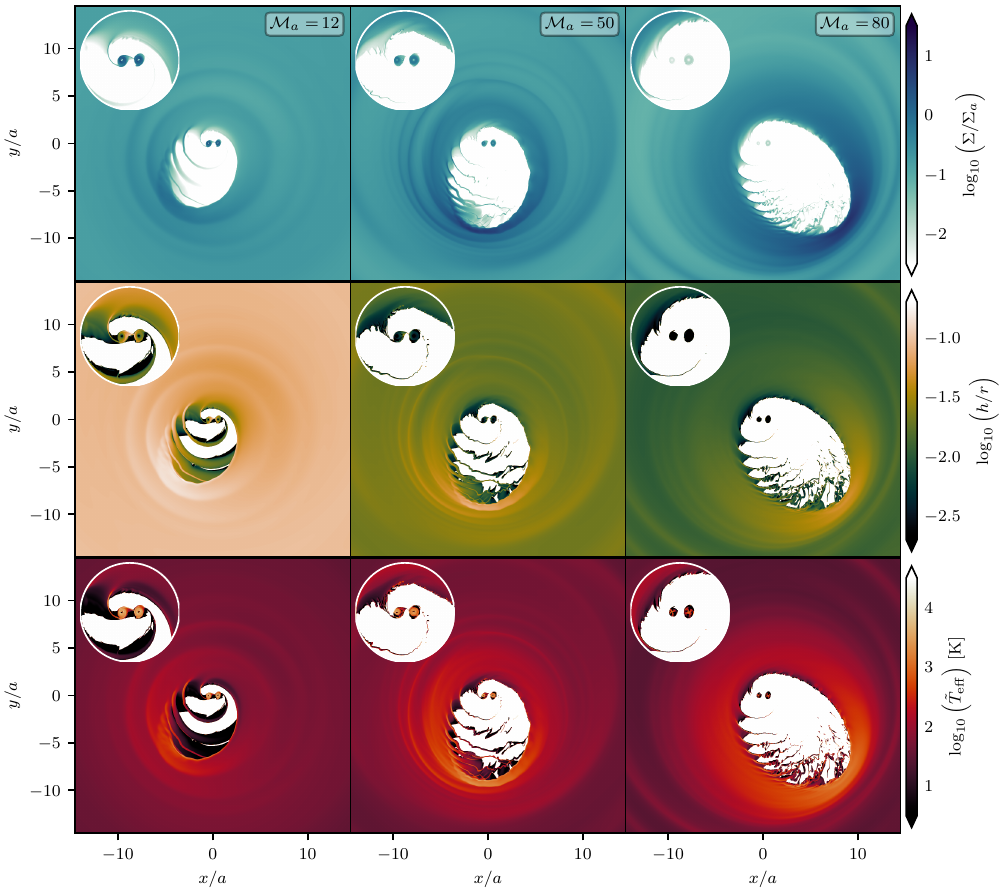}
    \vspace{-8pt}
    \caption{
    Snapshots after evolution for $\sim 3 t_\nu$ of surface density, aspect-ratio, and normalized effective temperature ($\tilde T_\eff$, corresponding to a remapping to an Eddington-accreting system described in Section~\ref{S:emission}) for each Mach number.
    Zoom-ins of the inner-most $3a$ of each panel are shown in the upper-left insets.
    }
    \label{fig:snapshots}
\end{figure*}

For the described heating and cooling procedures, these profiles ought to yield steady-state solutions around a single point mass.
We test this by initializing a $\Mach_a = 10$ disk around a $M = 100 \Msun$ sink region 
with an initial cavity of size $3a$.
The length $a$, here, is an arbitrary normalization and is simply taken to be large compared to the associated Schwarzschild radius. 
We set $r_{\rm sink} = a$ and evolve the disk for a viscous time at $10r_{\rm sink}$ for two values of the characteristic sink rate $s = \{20 t_\nu^{-1}, 40 t_\nu^{-1}\}$, where
\begin{equation}
    t_\nu \equiv \frac{2}{3}\frac{a^2}{\nu(a)} = \frac{a^2 \Omega_b}{\alpha \gamma} \frac{\Sigma_a}{P_a}
\end{equation}
is the viscous time at $r=a$ (c.f., \citealt{Dempsey2020}; and see Appendix~\ref{app:energy-sink} for more detailed discussion).
The associated solutions are illustrated in Figure~\ref{fig:single-profiles} as azimuthally-averaged radial profiles of the surface density, pressure, and rotational velocity, respectively.
The thick grey curves show the analytic solution, and the light-green line the initial configuration.
We observe that both solutions converge to the analytic disk profiles.
We adopt $s = 40 t_\nu^{-1}$ for all binary simulations presented below.

%
% ======================================================================
\section{Simulation results} \label{S:results}

We consider a binary of total mass $M = 10^6 \Msun$ with separation $a = 0.001 \,\unit{pc}$, corresponding to an orbital period $P_b \simeq 3 \, \unit{yr}$.
We place the binary in a box of total side length $50a$ with fiducial cartesian grid resolution $\Delta x = 0.015a$. We also run higher resolution segments, with $\Delta x = 0.0075a\ \textrm{and}\ 0.005a$, for convergence testing and some analyses, as specified below.
We explore solutions for three values of the Mach number, $\Mach_a = \{12, 50, 80 \}$, but because we fix $\alpha = 0.1$, this yields associated viscous times of $t_\nu = \{1.5 \times 10^2, 2.7 \times 10^3, 6.8 \times 10^3\}$ orbits, respectively.
Because of the very strong dependence of Eq.~\eqref{eq:mdot} on Mach number, the corresponding effective Eddington fractions are $\feff \equiv \Mdot_\infty / \Mdot_\edd \simeq \{1.3 \times 10^5, 10^2, 10 \}$, where $\Mdot_\edd$ is the Eddington accretion rate at 10\% radiative efficiency.\footnote{We comment that our goal is not to characterize the physics of super-Eddington accretion, but rather to document the relevant changes in binary accretion dynamics as one increases $\Mach_a$ and approaches the sub-Eddington limit.}
Lastly, in order to minimize artifacts from a violent startup-transient, we fix our coordinate center to the binary center of mass and adiabatically introduce the secondary component \citep[similar to the procedure in][]{Duffell2020} into the self-similar disk model by linearly increasing $\log_{10}(q)$ from an Earth-to-Sun mass ratio up to an equal mass binary over 2000 orbits.

Figure~\ref{fig:snapshots} shows snapshots of the solution surface density, aspect-ratio, and re-mapped effective temperature $\tilde T_\eff$ (described in Section~\ref{S:emission}) for the three fiducial runs evolved for $\sim 3\, t_\nu$ (corresponding to approximately 500, 10,000, and 20,000 orbits, respectively) at the fiducial resolution.
The top left of each panel also includes a zoom-in of the inner-most $3a$ of each snapshot.
In the surface density plots, we see that the cavity becomes increasingly elongated with increasing $\Mach_a$.
This is consistent with previously documented behaviors in circular, equal-mass binaries: that cavity semi-major axis grows with decreasing kinematic viscosity, and that cavity eccentricity grows with increasing Mach Number \citep[\eg][]{Munoz2020, Dittmann2022, Grcic:2026}.
Following the procedure of \citet{Dittmann2022}, we measure the cavity semi-major axis and eccentricity pairs as ($4.5a, 0.43$), ($5.6a, 0.46$), and ($7.2a, 0.55$) in order of increasing Mach number.
As $\Mach_a$ is increased, the rejected accretion streams propagating from the binary towards the cavity apocenter also become increasingly turbulent.
We posit that this results from an increasing susceptibility to Kelvin-Helmholtz (KH) roll-up due to (i) a decreased viscous damping of destabilizing modes, (ii) the increased filamentation of streams \citep[c.f.,][]{Tiede:2025} which shifts the characteristic disrupting modes to higher wavenumber, and (iii) the longer propagation time in more elongated cavities \citep{ChandrasekharBook61, Berlok19}.

One also observes that the density of the CBD inner edge, particularly at apocenter, grows with increasing $\Mach_a$, while the minidisk density falls.
This is consistent with increased accretion suppression with growing Mach number, such that
mass instead accumulates in a growing---in density and radial size---annulus at the cavity edge.
Here we comment that accretion suppression is principally a dynamical process governed by the removal of pressure support with growing Mach number, and is similarly observed at fixed viscous strengths \citep{Tiede2020, Dittmann2022, Tiede:2025}.
The disk also becomes thicker and hotter at cavity apocenter, generally coincident with the most prominent density enhancement.
We discuss in more detail in the following sections.

\begin{figure*}[t!]
    \centering
    \includegraphics{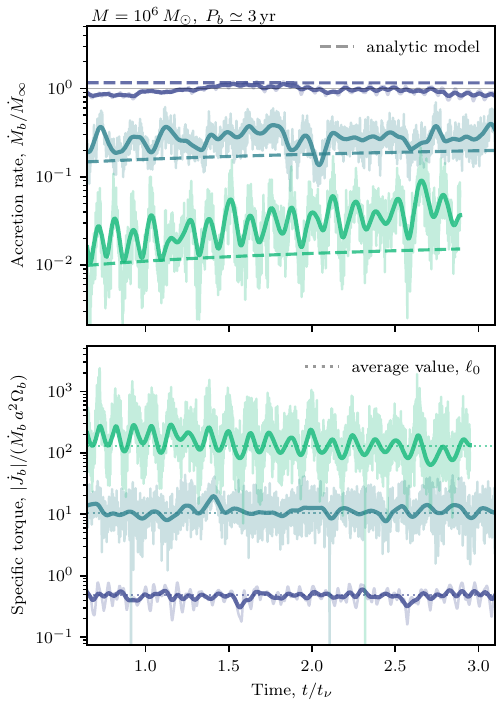}
    \includegraphics{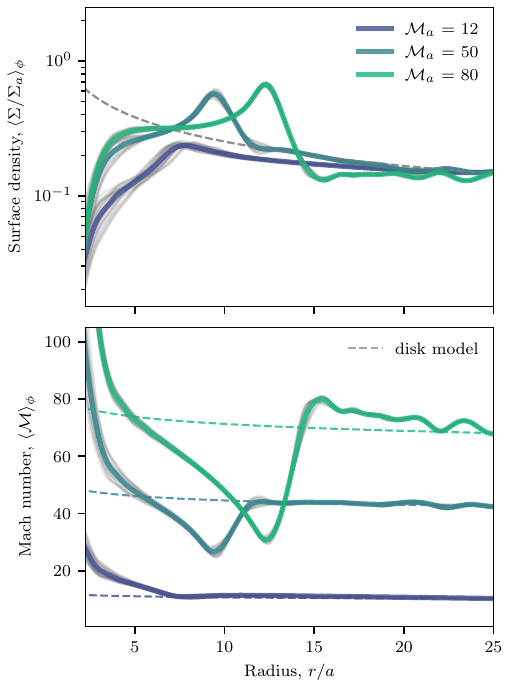}
    \caption{
    (Left) Timeseries of the binary accretion rate and specific torque magnitude from each of the three fiducial runs (note that the $\Mach_a = 12$ run corresponds to binary outspiral, while the higher Mach runs correspond to binary inspiral). 
    The dashed-lines in the top panel illustrate the analytic model for $\Mdot_b(t)$ computed from the empirical measurement of $\ell_0$.
    The dotted lines in the bottom panel indicate the measured magnitude of the torque parameter $\ell_0$ from an average over two viscous times.
    (Right) Azimuthally-averaged surface density and mass-weighted Mach number profiles. Individual profiles are shown by light grey lines and the time-averaged values over 500 orbits as solid, colored curves. Dashed lines indicate the reference disk model for each run. 
    }
    \label{fig:non-accrete-phase}
\end{figure*}
%

%
% ------------------------------------------------------
\subsection{Marginal- and non-accreting phases} \label{s:accretion-rates}

During each simulation, we track the total binary accretion rate through the sink source terms, $\Mdot_b(t)$, and the total torque, $\dot J_b(t)$, exerted on the binary by all cells in the domain.\footnote{We additionally compute and include the accretion torque, but this component is sub-dominant, especially in the limit of interest where $\Mdot_b < \Mdot_\infty$.}
For the time-evolution of the binary accretion rate,
\citet{Tiede:2025} found that a one parameter analytic model characterized by the \emph{torque parameter}, $\ell_0 \equiv \langle \dot J_b \rangle / \langle \Mdot_b a^2 \Omega_b \rangle$ (where $\langle \cdot \rangle$ denotes an average over a viscous time) very accurately described $\Mdot_b(t)$ for high Mach number configurations \citep[see also][]{Rafikov2016, SBCodeComp:2024}.
For the disk solutions considered here, with $\nu(r) \propto r^{3/5}$, this takes the form
\begin{align}
    \Mdot_b(t) = \frac{\Mdot_\infty}{1 - \ell_0\, (t / t_\nu)^{-5/14}} \ .
 \label{eq:mdot_t}
\end{align}
We present a framework for the physical interpretation of $\ell_0$ in Appendix~\ref{app:1d-picture} and a more detailed derivation of Eq.~\eqref{eq:mdot_t} in Appendix~\ref{app:time-evolution-model}.
We consider a system to be in steady-state when $\langle \Mdot_b \rangle \simeq \Mdot_\infty$, but note that because of the weak scaling with $t$ when $-\ell_0 \gg 1$, this can require hundreds to millions of viscous times. 
In this limit, $\Mdot_b \ll \Mdot_\infty$, and we refer to the binary as interacting with the disk in a \emph{non-accreting phase}.

The top-left panel of Figure~\ref{fig:non-accrete-phase} shows the time evolution of the binary accretion rate after the start-up transient for each of the three fiducial runs in time units of $t_\nu$.
The solid lines illustrate a running average over 2\% of a viscous time, and the faint curves, the un-smoothed timeseries.
The dashed lines of each color indicate the analytic evolution, Eq.~\eqref{eq:mdot_t}, estimated from an empirical determination of $\ell_0$.
Timeseries of the binary torque per unit accreted mass for determining $\ell_0$ are shown in the bottom-left panel of Figure~\ref{fig:non-accrete-phase}.
We refer to this as the specific torque, and
the average value is noted by a horizontal dotted line.
We see that, consistent with isothermal solutions, while the warm $\Mach_a=12$ system equilibrates rapidly, accretion is decreasingly efficient with increasing Mach number.
For $\Mach_a = 80$, this configuration yields a non-accreting phase, while the $\Mach_a = 50$ system can be characterized as marginally-accreting.

Unlike isothermal solutions, however, the average time evolution of the accretion rate is not precisely described by the analytic model.
While the approximation remains generally good for the steady-state and marginally-accreting solutions, it appears that $\Mdot$ evolves more quickly than $\sim t^{5/14}$ in the non-accreting system.
A primary assumption in Eq.~\eqref{eq:mdot_t}, demonstrated for isothermal solutions in \citet{Tiede:2025}, is that $\ell_0$ is a fundamental property of the system and does not change with time (or, equivalently, $\Mdot$).
However, once one includes an energy conservation equation,
the viscous dissipation rate, temperature, and scale height of the inner annuli all gradually grow during a non-accretion phase as mass slowly accumulates in the inner-disk.
For the two-region model of an inner-disk that has relaxed to the binary torque and an outer disk that retains its self-similar initial condition (see Appendix~\ref{app:time-evolution-model}), one can show that the dissipation rate of the relaxed inner-disk grows proportional to $\Mdot(t) \sim t^{5/14}$.
The associated scale-height, then, grows as a very weak function of time, $h \propto \Mdot^{1/5} \sim t^{1/14}$.
Yet, because $\ell_0$ varies strongly with $h$ (or equivalently, $\Mach_a$), even small changes to the inner scale-height induce a relevant effect.
We posit that this manifests as the modest secular evolution of $\dot J_b$ over the course of the $\Mach_a = 80$ simulation and accounts for the associated slightly faster than $t^{5/14}$ growth of $\Mdot_b$.
We explore further the degree to which $\ell_0$ varies as a function of accretion rate in Section~\ref{s:resolution}.

%
% ------------------------------------------------------
\subsection{Cold stream formation} \label{s:scale-height}

Despite the slight deviation from Eq.~\eqref{eq:mdot_t}, the magnitude of $\ell_0$ and weak growth of $\Mdot_b(t)$ still result in long-lived non-accretion phases for high Mach number systems.
This implies that, despite possible dynamical heating of the inner CBD, the material at stream launching remains cold.
From Figure~\ref{fig:snapshots}, one sees that streams are formed near cavity pericenter at $r \approx 1.5\, a$, while the density enhancement and dynamical heating effects (for the non-accreting runs) are most prominent at cavity apocenter.
These features are verified in radial profiles of the CBD surface density and Mach number in the right-panels of Figure~\ref{fig:non-accrete-phase}.
The light grey curves illustrate mass-weighted azimuthal averages from individual simulation snapshots and the colored curves depict the time-average over 500 binary orbits.
Dashed lines show the reference disk model of Section~\ref{S:disk-model}.
Each model generally maintains the analytic, single point mass solution at large radii, but as one approaches the binary, the $\Mach$-profiles show that the CBD is heated above the single-BH value at its far edge (cavity apocenter) and cooled below this reference value at its near edge (cavity pericenter; compare with Figure~\ref{fig:snapshots}).
The far side of the cavity is heated by a combination of the density pile-up and heating from stream generation that is advected along the cavity from its pericenter to apocenter.
These warm elements at cavity apocenter then cool efficiently on orbital approach to cavity pericenter, resulting in cold stream formation at local Mach numbers generally larger than $\Mach_a$.
We note that this cooling is dominated, not by radiative losses, but by the expansion of the flow as fluid elements return to cavity pericenter on eccentric orbits (see Appendix~\ref{app:compression-cooling}).
As a result, cold streams do not retain enough internal support to resist tidal compactification by the binary potential \citep{Dorazio2016, Tiede2022}, and one recovers long-lived non-accretion phases in the high-Mach limit.

%
% ------------------------------------------------------
\subsection{Numerical convergence and the runaway non-accretion problem} \label{s:resolution}
\begin{figure}[t!]
    \centering
    \includegraphics{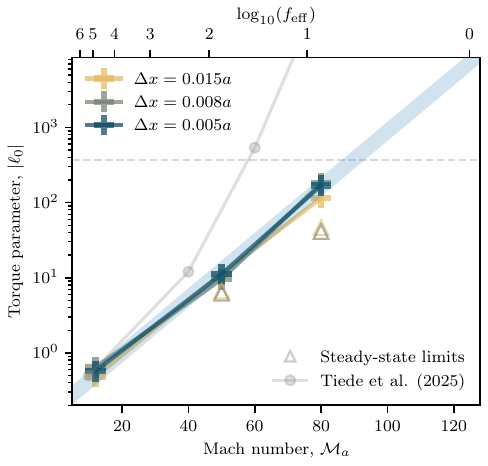}
    \caption{
    Magnitude of $\ell_0$ as a function of Mach number and resolution. 
    The top axis additionally illustrates the effective Eddington fractions associated with the disk model of Section~\ref{S:disk-model}.
    Crosses indicate runs initialized with no angular momentum current (Eqs.~\ref{eq:sigma_disk}~\&~\ref{eq:pressure_disk}), triangles indicate the determined steady-state limit by guessing the angular momentum current, and different colors correspond to increasing grid resolutions.
    The light blue band is an approximate exponential fit to the data.
    The grey curve illustrates the equivalent isothermal solutions, and the dashed-horizontal line depicts the critical value above which a binary will merge before it can ``turn-on'' (see Section~\ref{s:merger-timescales}).
    }
    \label{fig:resolution-test}
\end{figure}

\citet{Tiede:2025} demonstrated that $\ell_0$, and the associated accretion suppression, $\dot M_b / \Mdot_\infty \sim (-\ell_0)^{-1}$, grow exponentially with Mach number in isothermal solutions.
As a result, as one approached effective Eddington fractions $\feff \rightarrow 1$, the time required for an accreting binary to reach a steady-state rapidly exceeded not only the binary lifetime, but also a Hubble time.
This behavior was also magnified with increasing resolution.
We refer to this as the \emph{runaway non-accretion problem}.
To characterize how the problem changes with a more complete treatment of the energy equation, we plot the magnitude of $\ell_0$ from our simulations as a function of Mach number in Figure~\ref{fig:resolution-test}.
We also explore the degree to which our simulations accurately resolve stream formation and mass transfer across the binary cavity by ``upsampling'' the resolution by factors of 2 and 3, and measuring $\ell_0$ over an additional 1000 and 300 orbits, respectively.
We include as a grey curve the highest resolution results from the comparable isothermal simulations.

We find that our solutions are generally converged, with only mild variation from increasing resolution in the $\Mach_a = 80$ solutions.
We observe that, compared to isothermal configurations, the full treatment of the energy equation results in a more moderate exponential growth of $\ell_0$.
We compare our data with the function, $\propto e^{0.085 \Mach_a}$, plotted as a light blue band.
For comparison, the exponential growth in isothermal solutions was approximated as $\ln(\ell_0) \propto 0.21\, \Mach_a$.
However, we compute in Section~\ref{s:merger-timescales}---and illustrate as a horizontal, dashed line---the critical torque parameter above which a binary will merge before it can achieve an accretion rate 0.1$\dot M_\infty$.
From our reference fitting function, we estimate that this transition occurs above a Mach number of 90, and that as one approaches the limit $\feff \rightarrow 1$ (see e.g., the mapping on the top axis), binaries accreting from gas-pressure dominated disks will still generally merge during a non-accreting phase.

However, as noted in Section~\ref{s:accretion-rates}, as mass accumulates in the inner-disk during a non-accreting phase, the scale height of the disk will grow slowly in time.
We estimate the magnitude of this back-reaction on $\ell_0$ by guessing the steady-state angular momentum current (following the procedure of \citealt{MML17}, \citealt{MML19}) and initializing the disk using the corresponding steady-state profile (see Appendix~\ref{app:time-evolution-model}).
For each of the higher Mach numbers, we find setups that settle to accretion rates within 20\% of $\Mdot_\infty$ for $\sim 1000$ orbits at the fiducial resolution.
We increase the resolution by a factor of 2 and average over an additional $\sim 300$ orbits.
The associated measurements of $\ell_0$ are illustrated with triangles in Figure~\ref{fig:resolution-test}.
The torque parameter only changes slightly during the progression to steady-state for $\Mach_a = 50$, but it decreases by a factor of $2-3$ between the non-accreting and approximate--steady-state solutions for $\Mach_a = 80$.
These steady-state measurements act as lower-limits to $-\ell_0$ when estimating the time evolution from Eq.~\eqref{eq:mdot_t}, and their increasing trend with $\Mach_a$ suggests they may still approach the critical value as $f_\eff \rightarrow 1$.
We give more direct estimates for which binaries can and cannot ``turn on '' in Section~\ref{s:merger-timescales} below.

%
% ======================================================================
\section{Electromagnetic properties of truncated accretion} \label{S:emission}

\begin{figure}[t!]
    \centering
    \includegraphics{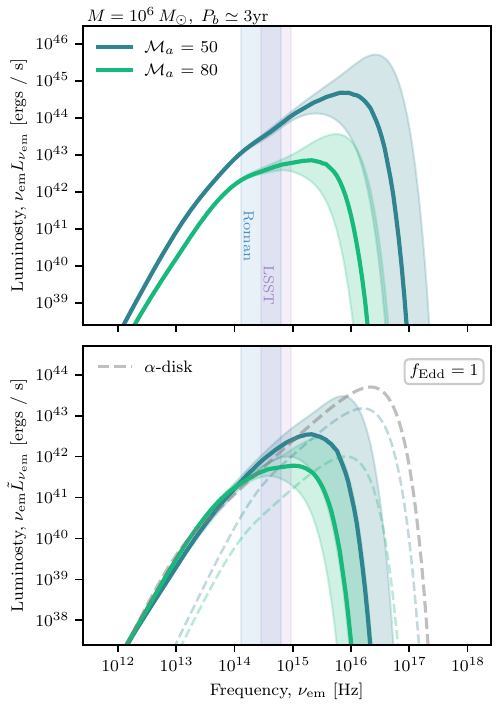}
    \caption{
    Average SEDs with 1$\sigma$ variability regions computed directly from the simulation data (top), and from the re-mapping to $f_\edd = 1$ (Eq.~\ref{eq:remapping}; bottom). 
    Vertical bands show LSST and Roman observing frequencies and the grey dashed line illustrates the SED of the equivalent $\alpha$-disk around a single BH. 
    Dashed-colored lines show possible sub-grid emission associated to an $\alpha$-disk solution around each black hole at the measured, respective, accretion rates.
    }
    \label{fig:seds}
\end{figure}

By evolving an energy equation and tracking the disk temperature, we can directly compute the thermal emission generated by the accretion flow.
In post-processing, we treat each cell as having a grey atmosphere and compute the effective temperature of each cell
\begin{align}
    T_\eff^4 = \frac{4}{3}\frac{T^4}{\tau(\rho, T)},
\end{align}
for optical depth $\tau$ and disk density $\rho = \Sigma / (2h)$.
We compute both the optical depth to scattering, $\tau_{\rm es} = \kappa_{\rm es} \Sigma$ and absorption, $\tau_{\rm abs} \approx \kappa_{\rm abs} \Sigma$.
We estimate the latter through Kramer's opacity law, $\kappa_{\rm abs} = 5 \times 10^{24} \rho\, T^{-7/2}\ \unit{cm}^2\ \unit{g}^{-1}$.
We verify that electron scattering opacity always dominates our solution in the disk and approximate the total optical depth as $\tau = \tau_{\rm es} + \tau_{\rm abs}$.
However, because many cells in the circumbinary cavity become extremely diffuse, we additionally compute their effective optical depth $\tau_\eff = \sqrt{\tau_{\rm abs}(\tau_{\rm es} + \tau_{\rm abs})}$ \citep{RybickiLightman:RPA}.
To compute spectral energy distributions (SEDs) from our solutions, for all cells that are thermalized with $\tau_\eff > 1$, we sum their differential specific luminosity $dL_\nuem = \pi \mathcal{B}_\nuem [T_\eff]\, d\nuem\, dA$ for each frequency $\nuem$ over cell areas $dA$; with $\mathcal{B}_\nuem$ Planck's function.
We compute the averaged SEDs from snapshots of the highest resolution, $\Delta x  = 0.005\,a$, simulations taken 10 times per orbit for approximately 50 orbits.
Because we are interested in the characteristics of non-accreting (or marginally-accreting) phases, we restrict our emission analysis to the Mach 50 and 80 runs.
The SEDs computed directly from each simulation are illustrated in the top panel of Figure~\ref{fig:seds}.
The solid lines show the average luminosity at each frequency, while the shaded regions indicate the associated one-standard-deviation variability region. 
The SEDs appear generally consistent with a truncated $\alpha$-disk, but with growing variability with increasing frequency due to the periodic passage of material near each binary component during stream formation.

Recall that the effective Eddington fraction of the Mach 50 and 80 simulations were $\feff \approx$ 100 and 10, respectively.
In the bottom panel of Figure~\ref{fig:seds}, we plot our SEDs after rescaling our fluid quantities \citep[following the general strategy of][]{Westernacher-Schneider:2022} to a reference system that is accreting at its Eddington rate, $f_\edd = 1$.\footnote{Note, however, that \citet{Westernacher-Schneider:2022} only re-mapped $T_\eff$; but because we additionally compute $\tau_\eff$, we apply the re-mapping to all relevant fluid quantities.}
In particular, because Equations~\eqref{eq:sigma_disk}~\&~\eqref{eq:pressure_disk} are self-similar in $\dot M_\infty$, we perform the map
\begin{equation}
 \begin{aligned}
    \Sigma &\to \tilde \Sigma = \Sigma\, (f_\edd / \feff)^{3/5} \\
    P &\to \tilde P = P\, (f_\edd / \feff)
 \end{aligned}
 \label{eq:remapping}
\end{equation}
in order to compute the re-scaled values, $\tilde T_\eff$ and  $\tilde \tau_\eff$.
Here we make a distinction between $f_\edd$, when referencing the self-similar disk solutions, and $\feff$ when referring specifically to our simulation setups.
We compare the resultant SEDs with that of a standard $\alpha$-disk around a $10^6 \Msun$ black hole with $f_\edd = 1$ and $\alpha=0.1$, shown as a grey dashed curve.
We see that the non-accreting (and marginally-accreting) solutions, on average, have much lower bolometric luminosities than the equivalent single black hole.
Despite this overall reduction, each system remains relatively luminous at optical and near infrared (NIR) frequencies corresponding to emission from the heated region at the far end of the binary cavity (see, e.g, bottom row of Figure~\ref{fig:snapshots} and right panels of Figure~\ref{fig:non-accrete-phase}).
We plot as faint vertical bands the range of observing frequencies for LSST \citep[][purple]{LSST:ScienceBook:2009} and the Nancy Grace Roman Space Telescope \citep[][light blue]{Roman2015}.
For the non-accreting solution, the system remains comparably luminous to a single, accreting SMBH at most of the Roman observing frequencies and the low-end of the optical LSST bands, but it starts to lose relevant optical luminosity towards the bluer portion of LSST.
Both solutions also posses variability in both the Roman and LSST bands---for example at the $\approx 0.2\,\unit{dex}$ level in the overlap region around $\nuem = 5 \times 10^{14}\, \unit{Hz}$---such that periodicity would likely remain detectable in observed lightcurves.
We make more explicit detectability estimates in Section~\ref{s:detectability-estimates}.

We additionally comment that these measurements, especially the non-accreting solution, represent an early evolutionary phase of the accreting binary. 
As the inner CBD relaxes to the binary-imposed angular momentum current, the local dissipation increases compared to the standard, single-BH solution. 
In the time-dependent model motivated in Section~\ref{s:accretion-rates} (and described in more detail in Appendix~\ref{app:time-evolution-model}), the effective temperature of the relaxed inner-disk grows, albeit slowly, as $T_\eff \propto \Mdot_b^{1/4} \sim t^{5/56}$. 
We therefore expect the SEDs in Figure~\ref{fig:seds} to evolve to larger optical/NIR luminosities as the inner-disk accumulates mass and transmits a growing net angular momentum current \citep[see also,][]{Kocsis+2012a, Kocsis+2012b, Rafikov2013}. 
The spectral peak should also gradually increase in amplitude and shift blue-wards as $T_\eff$ rises.
As a result, however, those annuli contributing most strongly in the optical/NIR (i.e., those at $r_{\rm em}$ satisfying $k_b T_\eff(r_{\rm em}) \sim \rm{h}\nuem$) will shift to larger radii where dynamical processes sourcing variability at the binary orbital period may be less prominent.

Lastly, because we do not resolve down to horizon scales, we include as colored, dashed lines the omitted emission---implied by the average accretion rates of Figure~\ref{fig:non-accrete-phase}---integrated in radius from the Schwarzschild inner-most stable circular orbit of each black hole out to the sink radius.
This assumes that the minidisks remain radiatively efficient, but as $\Mdot_b / \Mdot_\infty$ falls, the circum-single flows become susceptible to transition into radiatively inefficient solutions with yet lower luminosities and different spectral characteristics (\citealt{NarayanYi:ADAF:1994}; and see \citealt{Tiede:badaf:2025} for a discussion of binaries in such multi-component flows.)
This emission is, expectedly, significant for the marginally-accreting case, but is only minor in the regions of interest for the non-accreting solution.
We discuss more detailed implications for electromagnetic searches for SMBHBs in the following section.

%
% ======================================================================
\section{Discussion} \label{S:discussion}
%

%
% ------------------------------------------------------
\subsection{Numerical approximation and caveats} \label{s:caveats}

We have demonstrated that long-lived binary non-accretion phases extend to more detailed treatments of the energy equation, particularly because stream formation occurs from comparatively cold material.
We have limited our inquiry, though, to classical $\alpha$-disk models dominated by thermal pressure, and a number of processes may yet alter the thermodynamic structure of the inner-disk, stream formation, and the existence of non-accretion phases.
In particular, as a binary shrinks its orbit towards merger, radiation will at some point begin to dominate the energy content of the inner-CBD and may qualitatively change this picture \citep[c.f.,][]{Cocchiararo:2025, Tiwari:2025}. 
We have also neglected the direct inclusion of magnetic fields, which can offer an additional source of gas support and access to alternate accretion modes \citep[i.e.,][]{Most:MHD:2024, Hopkins:MAD-Quasars:2024, Wang:BMAD:2025}.
The presence of magnetically driven turbulence as well as tension and pressure from ordered magnetic fields will impact stream generation and the dynamics of mass transfer onto the binary \citep[e.g.,][]{Shi+2012, Noble2021, EnnoggiKrolik+2025}.
We have additionally modeled the flow as a vertically averaged, two-dimensional flow, whereas, despite the increasingly thin nature of the solutions, the inclusion of a vertical axis may alter the cavity structure, stream geometry, and gas capture dynamics.
The presence of orbital eccentricity \citep[e.g.,][]{Zrake2020, Dorazio2021, Dittmann:2026} or an unequal mass binary \citep[e.g.,][]{Duffell2020, Siwek2023, Dittmann:q-mach:2024} may similarly alter this mass transfer process.

We have additionally assumed in our cooling prescription that the solution is everywhere, locally characterized as a blackbody.
This approximation, though, does not generally hold for the lowest-density regions of the accretion flow (i.e., the cavity), and in future work, we plan to modify the cooling function when $\tau_\eff \lesssim 1$.
Similarly, in the blackbody cooling, we have adopted a constant electron-scattering opacity when computing the radiative energy losses; and while we have checked that this approximation is generally appropriate for the code densities and temperatures (corresponding with super-Eddington effective accretion rates), a more detailed treatment of the gas opacities may also alter the disk structure and emission in setups that physically correspond to sub-Eddington accretion.
Therefore, our results should be viewed as isolating the effect of evolving an energy equation relative to simpler thermodynamic closures, rather than as a complete model of binary accretion.

%
% ------------------------------------------------------
\subsection{Physical interpretation and application}
\label{s:interpretation-application}

As demonstrated, the critical parameter determining the time evolution, steady-state structure, and characteristic timescales of an accreting binary is the torque parameter, $\ell_0$.
Physically, $\ell_0$ reflects the efficiency with which mass pulled into accretion streams is deposited onto binary components, where $-1 / \ell_0$ is the \emph{stream efficiency} (\citealt{Tiede:2025}; and see Appendix~\ref{app:1d-picture} for a more detailed discussion). 

The existence of long-lived non-accreting phases and the exponential growth of the torque parameter (e.g. Figure~\ref{fig:resolution-test}) occur because of an increasingly inefficient hydrodynamic coupling between the binary and the inner-disk as pressure support is decreased. 
These effects are similarly observed in simulations of cold finite-disks in 3D \citep{Ragusa+2016} and 2D circumbinary rings \citep{Betancourt:2026} indicating that the core physics of mass transfer are robust to initial configuration.
Thus, while our modeling is in reference to an infinite disk, we posit that the determined values of $\ell_0$ and associated evolutionary timescales are applicable to more complex or realistic initial conditions.
Namely, unless the initial mass distribution is highly concentrated to scales of the binary separation, a generic initial disk will spread viscously until its inner edge begins to interact with the binary potential. 
If the disk is relatively cold, the accretion rate will be suppressed until the disk relaxes to meet one of three criteria:
\begin{enumerate}
    \item It viscously delivers sufficient material to the inner-disk to compensate for the low stream efficiency and enable a meaningful mass accretion rate.
    \item The stream efficiency grows with increasing inner-disk temperature and scale-height as material piles up, sufficient to turn-on accretion.
    \item The accumulated inner-disk mass grows so large that it becomes gravitationally unstable.
\end{enumerate}
Option (1) generally corresponds to the system achieving steady-state and requires the disk to viscously relax out to the radius $r_0 \equiv \ell_0^2 a$, where the specific angular momentum is that associated with the torque parameter, $\ell_0 a^2 \Omega_b$.
The typical time for this is the viscous time at $r_0$, which corresponds to achieving $\Mdot_b / \Mdot_\infty = 1/2$ in Eq.~\eqref{eq:mdot_t}.
Depending on the value of $\ell_0$, though, option (2) and (3) may occur before reaching this steady-state.

For option (2), the key point is that $\ell_0$ (and the stream efficiency) depends very steeply on the scale-height of the inner disk. 
As a result, there is a critical initial $\Mach_a$ (i.e., in the steady disk with no angular momentum current) above which the inner-disk will become hot enough to efficiently feed the binary before the system has viscously relaxed to $r_0$. 
Using the functional form for $\ell_0(\Mach_a)$ from Section~\ref{s:resolution}, and noting that $h(r,\ell_0) \propto (-\ell_0)^{1/5}$ for $-\ell_0 \gg 1$ in our disk
model (see Eq.~\ref{eq:steady-height}), we find that the inner-disk reaches $h/r \sim 0.1$ before option (1) is satisfied only for $-\ell_0 \gtrsim 10^6$. 
This corresponds to initial configurations with $\Mach_a \gtrsim 200$, or equivalently $f_\eff \lesssim 0.1$ for fiducial system parameters.

To evaluate when option (3) may occur, we compute the Toomre parameter of the inner-disk in our model, 
\begin{align}
    Q_{\rm in}(\ell_0) = \frac{c_s(r_{\rm in}, \ell_0)\, \Omega(r_{\rm in})}{\pi G\, \Sigma(r_{\rm in}, \ell_0)} \simeq 0.002\, \Mach_a^2\, \ell_0^{-2/5} \nonumber
 % \label{eq:toomre}
\end{align}
for a $10^6 \Msun$ binary with a 1 year period and $-\ell_0 \gg 1$.
By again assuming the functional form of $\ell_0(\Mach_a)$, we find that $Q_{\rm in}$ falls below one---and the disk becomes gravitationally unstable due to inner mass pile-up---when initial $\Mach_a \gtrsim 130$, corresponding to $-\ell_0 \gtrsim 10^5$.

Therefore, for initial Mach numbers exceeding $\approx 130$, the inner-binary may start accreting efficiently once the inner-disk begins to destabilize under its own self-gravity and generates additional heat and pressure support from the onset of gravitational instability \citep{Gammie:2001, Franchini2021}.
At even higher Mach numbers, continued mass accumulation and associated increases in the inner-disk temperature, scale height, and stream efficiency can trigger binary accretion before the system reaches the viscously relaxed steady-state.
Each of option (2) and (3), however, require (in our phenomenological model) very large values of $-\ell_0$, which will generally drive the binary to merge before it can turn on.
We demonstrate this in the following section.

%
% ------------------------------------------------------
\subsection{Merger timescales and turn-on conditions} \label{s:merger-timescales}

We have established two primary regimes in the evolution of binaries accreting from high Mach number disks.
At early times, the binary interacts with a truncated CBD in a non-accreting phase, where $\dot m \equiv \Mdot_b / \Mdot_\infty \ll 1$.
At late times, the binary achieves a steady-state where $\dot m \rightarrow 1$.
We refer to the transition between the two as the binary ``turning-on''.\footnote{This turn-on roughly corresponds to the phase in which the average emission transitions from the truncated SEDs computed in Figure~\ref{fig:seds} to more standard AGN-like spectra.}
Whether a binary reaches steady-state or not---and how long it takes to merge---depends on if the required turn-on--time is long or short compared to the binary lifetime.

The characteristic merger time for a circular, equal mass binary accreting at $\dot m$ is
\begin{align}
    \frac{a}{\dot a} &= \frac{1}{8 \dot m } \left( \ell_0 - \frac{3}{8} \right)^{-1}  t_M \ ,
 \label{eq:adot}
\end{align}
where $t_M = M / \Mdot_\infty$ is the typical mass doubling time.
However, under the time evolution model of Eq.~\eqref{eq:mdot_t}, in a non-accreting limit with $-\ell_0 \gg 1$, the merger time becomes independent of the torque parameter,
\begin{equation}
    \frac{a}{\dot a} \approx \frac{t_M}{8}\, \left( \frac{t}{t_\nu} \right)^{-5/14} \ .
 \label{eq:adot_t}
\end{equation}
The inspiral rate, thus, grows over viscous times---in a manner that depends only on the disk model---as material accumulates in the inner-disk, exerts an increasingly large torque, and transmits a progressively growing amount of angular momentum through the disk.

To this point, though, we have approximated the accretion rate evolution under the assumption that $\ell_0$ is a fixed property of the system.
However, we have demonstrated that when one allows the scale-height of the inner-disk to adjust self-consistently, $\ell_0$ evolves slowly on the order of viscous times (e.g. Figure~\ref{fig:non-accrete-phase}~\&~\ref{fig:resolution-test}).
Assuming this evolution mirrors that of $\Mdot$, we model it phenomenologically as
\begin{align}
    \ell_0(t) = \ell_1 (t / t_\nu)^{-p} \ ,
 \label{eq:ell0_t}
\end{align}
where $\ell_1$ is the torque-parameter early in the evolution (at one viscous time), and generally corresponds to the values measured in our simulations (e.g., the crosses in Figure~\ref{fig:resolution-test}).
Inverting Eq.~\eqref{eq:mdot_t} for the turn-on time $t(\dot m)$, and equating it with the associated merger time through Eqs.~\eqref{eq:adot_t}~\&~\eqref{eq:ell0_t} in the limit that $-\ell_0 \gg 1$, defines a critical value for the torque parameter
\begin{align}
    |\ell_{1,\,\crit}| \approx 2^{-3 \xi} \, \dot m^{-1} \left( t_M / t_\nu \right)^{\xi} \ ,
 \label{eq:ell1-crit}
\end{align}
with $\xi = (14p + 5) / 19$,
above which a binary will merge before it can achieve some $\dot m$ and turn on.\footnote{The coefficients of $\xi$ come from our choice of disk model. In the more general parameterization of Appendix~\ref{app:time-evolution-model}, $\xi = (\lambda + p) / (\lambda + 1)$.}

The expression for $\ell_{1,\, \crit}$, however, depends on the local viscous time, and thus, changes as the binary orbit shrinks.
Because $t_\nu$ is an always decreasing function with decaying $a$, the absolute limit
occurs near the point where the binary decouples from its accretion disk due to runaway gravitational wave (GW) emission \citep[e.g.,][]{Armitage2002}.
The binary lifetime from GW evolution is $t_{\rm GW} = \mathcal{A}^{-1} a^4$, with  $A = 16 (GM)^3 / (5c^5)$ for an equal mass binary \citep{Peters1964}, and the critical separation below which a binary can never turn on occurs when $t_{\rm GW} = t(\dot m)$.
This occurs at
\begin{align}
    a_\crit \simeq 10^{-5} \ f_\edd^{-2/13} \alpha_{0.1}^{-4/13} M_6^{14/13} (\dot m \ell_1)^{70/(247\xi)} \ \unit{pc} \nonumber \ ,
\end{align}
where we've defined typical values $\alpha = 0.1 \alpha_{0.1}$ and $M / \Msun = 10^6 M_6$.
Choosing a value $p=1/4$, visually consistent with the data in Figure~\ref{fig:non-accrete-phase}, and
evaluating Eq.~\eqref{eq:ell1-crit} at $a_\crit$ yields
\begin{align}
    |\ell_{1,\,\crit}| \simeq \big( 360\, \dot m_{0.1}^{-1}\big)\,  f_\edd^{-17/138} \alpha_{0.1}^{136/345} M_6^{-289/690} \nonumber  \ ,
\end{align}
% \dd{do we want to bring this full circle with scenarios 2 and 3 above, where we also estimated a kind of critical $l_0$}
where we've additionally defined the turn on condition, $\dot m = 0.1 \dot m_{0.1}$.
This is multiple orders of magnitude smaller than the estimated torque parameter's at which the inner-disk becomes self-gravitating or heats up to $h/r\sim0.1$, reinforcing that the binary would merge before achieving scenario (2) or (3) in Section~\ref{s:interpretation-application} above.

The torque parameter values determined from our simulations (see e.g., Figure~\ref{fig:resolution-test}) begin to approach $\ell_{1,\,\crit}$ as we increase $\Mach_a$, and setups approaching the limit $\feff \rightarrow 1$ would likely merge before achieving $0.1 \Mdot_\edd$.\footnote{We comment that when $p = 0$, relevant to isothermal solutions, $\ell_{1,\,\crit} \rightarrow \ell_{0,\,\crit}  \simeq 80\, \dot m_{0.1}^{-1} f_\edd^{-5/69} \alpha_{0.1}^{16/69} M_6^{-17/69}$.}
We estimated in Section~\ref{s:resolution} that this transition occurs for $\Mach_a \gtrsim 90$, such that systems with effective Eddington fractions
\begin{align}
    f_\eff \lesssim 5.5\ \alpha_{0.1}^{1/2} \, M_6^{3/4}\, (a /\, 10^{-3}\unit{pc})^{-1/4} \nonumber
\end{align}
would, in our disk model, merge in a non-accreting phase.
We also comment, that any system with torque parameter much larger than $|\ell_{1,\,\crit}|$ may be more simply modeled as a "non-accreting" disk with $\Mdot = 0$ and steadily growing torque that is independent of the actual value of $\ell_0$ \citep[e.g.,][]{Rafikov2013}.

\begin{figure}[t!]
    \centering
    \includegraphics{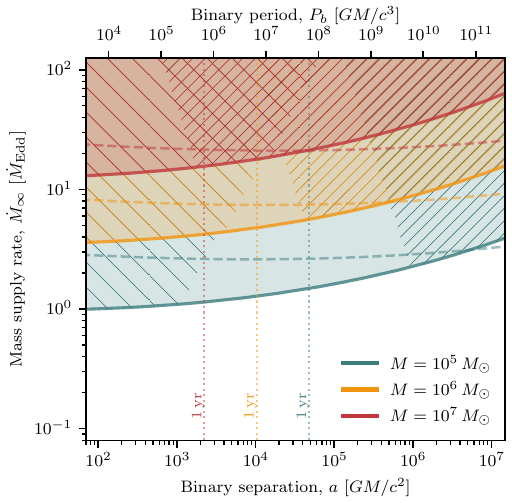}
    \caption{
    Regions of binary separation and supply rate where $t_{\rm merge}$ exceeds $t(0.1 \Mdot_\edd)$ for three values of the binary mass with $\alpha= 0.1$.
    Solid lines and accompanied shaded regions are for $p=1/4$, while dashed lines indicate the constant$-\ell_0$ limit, $p=0$.
    Binaries below each curve will merge before appearing similar to a standard AGN.
    Shaded-hatched regions of each color correspond to those where the disk model becomes radiation pressure dominated (left, backslash) or unstable to its own self-gravity (right, forward-slash).
    Vertical dotted lines show 1 year rest-frame binary periods for each mass.
    }
    \label{fig:turn-on-conditions}
\end{figure}
\begin{figure*}[t!]
    \centering
    \includegraphics{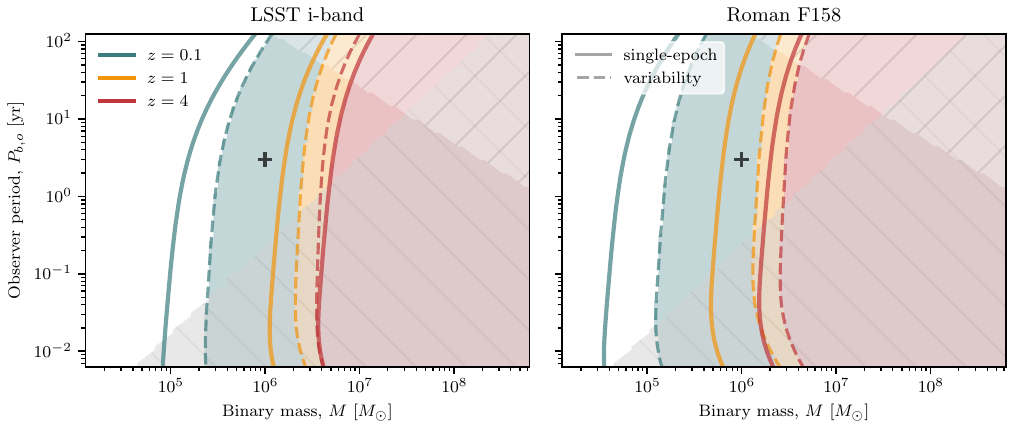}
    \caption{
    Estimate of detectability for truncated binary accretion that retains periodic variability from the inner-disk.
    Solid lines illustrate single-epoch flux detectability and dashed lines the possible detection of periodic variability.
    The black cross indicates the mass and period corresponding to our simulations.
    Hatched-shaded regions show where a standard $\alpha$-disk with $f_\edd = 1$ becomes approximately gravitationally unstable (top) and dominated by radiation pressure at $r \simeq 3 a$ (bottom).
    }
    \label{fig:detectability}
\end{figure*}

If a binary merges during a non-accreting phase,
because Eq.~\eqref{eq:adot_t} evolves explicitly with time, the characteristic merger time is no longer self-similar in $\dot m$.
It instead depends on the binary separation at the onset of accretion, and
integrating Eq.~\eqref{eq:adot_t} from $0.1a_{0.1}\,\unit{parsec}$ to merger gives a lifetime,
\begin{equation}
    t_{\rm merge} \simeq (0.4\, t_M) \,  f_\edd^{3/19} \alpha_{0.1}^{-4/19} M_6^{-1/19} a_{0.1}^{7/19} \nonumber \ ,
\end{equation}
where $t_M \approx \big(45\, \unit{Myr} \big)\, /\, f_\edd$ is the mass doubling time at 10\% radiative efficiency.
To visualize which binaries merge before they can turn on, we plot in Figure~\ref{fig:turn-on-conditions} those curves where $t_{\rm merge} = t(\dot m)$ for varying binary separations (in gravitational radii) and mass supply rates (in units of the Eddington rate), assuming our approximate fit, $\ln(\ell_1) = 0.085\, \Mach_a - 1.7$, from Figure~\ref{fig:resolution-test}.
Because we vary $\Mdot_\infty$, we choose as a turn-on condition $\Mdot_b = 0.1 \Mdot_\edd$.
Shaded regions illustrate those binaries that can turn-on before merger for $\alpha = 0.1$ and $p = 1/4$, and the dashed curves show how these regions change in the constant$-\ell_0$ limit, with $p=0$.
Hatched areas illustrate where the fiducial disk model becomes dominated by radiation pressure (left, backslash) or self-gravity (right, forward-slash) for each mass.
We find that all massive binaries require mass supply rates exceeding $\Mdot_\edd$ in order to turn on, but that those with periods of years and total mass $\lesssim 10^6 \Msun$ can turn on prior to merger.
However, all binaries with $ \gtrsim 10^7 \Msun$ in gas-pressure dominated disks will be driven to merger in a non-accreting phase.

For those binaries that do turn on at some time $t_{\rm on}$, and reach a steady-state, from Eq.~\eqref{eq:adot}, they will merge on a timescale $t_M / (8 \ell_0(t_{\rm on}))$. 
This equates to a ``minimum bright lifetime'' of approximately
\begin{align}
    0.15\, f_{\edd}^{-121/138} \,\big( \ell_0(t_{\rm on}) / \ell_{1,\,\crit} \big)^{-1}\,\unit{Myr} \nonumber
\end{align}
for fiducial values of $\alpha$, $M$, and $p=1/4$.
The quantity $\ell_0(t_{\rm on})$ can be estimated from the quasi--steady-state solutions (triangles) in Figure~\ref{fig:resolution-test}.

%
% ------------------------------------------------------
\subsection{Detectability of truncated binary accretion}
\label{s:detectability-estimates}

The SED results of Figure~\ref{fig:seds} are for a specific binary configuration, but a general detectability estimate for massive binaries interacting with a CBD in a non-accreting phase will vary with the binary mass, period, and redshift ($z$).
In order to estimate which set of systems may be identifiable in time-domain surveys, we compute both single-epoch and periodic detectability estimates for representative bands in the Legacy Survey of Space and Time (LSST) and the Roman Space Telescope.
We approximate the SED of a non-accreting binary analytically as a truncated $\alpha$-disk and compute the average observed flux density $F_\nuobs = (1+z)\, L_{\nuem}/(4\pi d_L^2)$ at observer frequency $\nuobs=\nuem\, (1+z)^{-1}$ and luminosity distance $d_L(z)$ in standard $\Lambda$CDM cosmology.
We convert this to an AB magnitude and compare against a single epoch detection threshold $m_{\rm lim}$, adopting $m_{\rm lim}=24$ for the LSST $i$-band and $m_{\rm lim}=26$ for the Roman F158 filter. 

To assess periodic detectability, we compute the characteristic fractional flux variability, $\Delta F_\nuobs / F_\nuobs$, from the $1\sigma$ deviations of our $\Mach_a = 80$ SED (i.e., the green shaded region in Figure~\ref{fig:seds}) and assume that lightcurves vary periodically at this amplitude.
We do not attribute any characteristic timescale to this variability.
In order to estimate across binary masses and periods, we also assume that the magnitude of the SED variability is a fixed function of the disk radius in gravitational radii, $r_g$, for our reference truncated $\alpha$-disk.
Because the effective temperature of an $\alpha$-disk scales as $T_\eff \propto r^{-3/4}$ away from the horizons, the characteristic radius emitting at $\nuem$  scales as $r_\nuem  \propto (M \nuem^4)^{-1/3}\, r_g$ as we vary the binary mass at fixed $f_\edd$.
We, therefore, attribute the $1\sigma$ variability of our fiducial $10^6 \Msun$ model to its characteristic emission radius and map this to equivalent values of $r_\nuem / r_g$ as we vary the binary mass and period.
In order to detect the associated flux variability, we require it to exceed the threshold
\begin{align}
    % \Delta F_\nuobs > \big( F_\nusen / F_\nuobs + 0.05 \big) F_\nuobs \nonumber \ ,
    %
    \Delta F_\nuobs > F_\nusen + 0.05\, F_\nuobs \nonumber \ ,
\end{align}
with $F_\nusen$ the survey flux sensitivity, and 5\% the assumed minimum detectable variation \citep{Kelley:LensingDetect:2021}.

We plot in Figure~\ref{fig:detectability} our approximations for both the single-epoch flux detectability limit (solid lines) and the variability detection limit (dashed lines) for relevant binary masses and periods at three redshifts.
We ignore limits from the associated survey lifetimes (about ten years for LSST and two years for the Roman HLTDS).
The black cross indicates the binary corresponding to our simulations, and color-shaded regions illustrate those binaries that are both bright and variable enough to be detected at a given redshift.
We conclude that, in our broad approximation, nearly all relevant massive binaries in the local universe ($z \lesssim 0.1$) interacting in non-accreting phases remain detectable in upcoming time domain surveys, while only those with total mass $\gtrsim 2 \times 10^6 \Msun$ remain detectable at $z \gtrsim 1$.
At larger binary masses, though, the hatched regions indicate where the underlying assumptions of an ionized disk dominated by thermal pressure break down because the surrounding disk becomes either unstable under it's own self-gravity (light-grey, top) or dominated by radiation pressure (dark-grey, bottom; \citealt{HKM09, Haiman2009:Erratum}).

We additionally, comment that at large redshift, optical filters begin to sample the UV turnover in the truncated rest-frame SED, where $\Delta F_\nuobs$ can exceed $F_\nuobs$.
In such cases, a binary may not be detectable as a steady source, but may appear (or disappear) in difference imaging over a binary period.
We plan to explore the full nature of the periodic variability from our simulations in follow up work.

%
% ------------------------------------------------------
\subsection{Ionizing fluxes and intermittent accretion episodes}

A significant consequence of the truncated, non-accreting phases considered here is the strong suppression of the high energy continuum emission compared to a standard single black hole $\alpha$-disk.
This implies that binaries in non-accreting phases, even if they remain detectable in optical/IR surveys, are intrinsically X-ray weak and inefficient at producing the photo-ionizing radiation required to sustain broad- and narrow-emission line regions typically observed in AGN.
In this sense, binaries interacting with truncated CBDs in a non-accreting phase may manifest as a form of weak emission-line quasars \citep{Fan:FirstWLQ:1999, Dioamond-Stanic:WLQs:2009, Plotkin:SDSS-WLQs:2010, Wu:XrayWeakQuasars:2011, Meusinger:WLQs:2014}.
This is highlighted by the bottom panel of Figure \ref{fig:seds} where the SED turnover, in this example, occurs near the Lyman limit. 
This turnover would result in diminished hydrogen recombination line strengths compared to standard, single-BH AGN, as well as a near complete suppression of high-ionization lines.
Systems with larger masses or longer orbital periods would also experience cutoffs at even lower frequencies and exhibit further attenuated ionization features.

Moreover, intermittent feeding of the minidisks---as indicated by the super-orbital-timescale oscillations between relative high- and low-accretion states in the high-$\Mach_a$ timeseries of Figure~\ref{fig:non-accrete-phase}---may also suggest that the ionizing photon flux can vary meaningfully with time. 
In particular, the hierarchy of relevant timescales may provide a plausible route to changing-look--like behavior in truncated binary accretion.
Namely, in a relatively high-accretion phase, the supply of material to the minidisks occurs on comparatively short dynamical times, while the persistence and decay of enhanced ionizing fluxes from this feeding is regulated by the viscous time in the minidisks (many binary orbits), and the duty cycle of strong overfeeding may be modulated on yet longer circumbinary timescales (e.g., by large-scale non-axisymmetric structures or cavity precession).
We suggest that truncated binary accretion may offer qualitative connections to both weak-line quasars and changing-look AGN through its impact on the ionizing continuum, but we defer a more quantitative analysis to future work.

%
% ======================================================
\vspace{-5pt}
\section{Conclusions} \label{S:conclusions}

We have presented hydrodynamics simulations of circumbinary accretion onto equal-mass SMBHBs in classically thin, gas-pressure dominated disks while evolving an energy equation that includes viscous and hydrodynamic heating coupled with radiative blackbody cooling. 
Our goal was to test whether the long-lived non-accretion phases found in locally isothermal calculations survive once the inner CBD is allowed to heat, cool, and self-consistently adjust its scale height.

Our principal result is that non-accretion phases persist for such thin disks when including a treatment of the energy equation.
Although mass pile-up and stream--cavity interactions heat the CBD near cavity apocenter, the gas that sources tidal streams originates from comparatively cold material near cavity pericenter. 
The streams therefore remain insufficiently pressure-supported to resist tidal compactification by the binary potential, and accretion onto the binary remains suppressed in the classically thin, high Mach number limit. 
We find that the inclusion of an energy equation moderates the growth of the torque parameter with increasing Mach number---relative to locally isothermal solutions---but does not completely assuage the runaway non-accretion problem.
Thus, as the effective Eddington fraction of our gas-pressure dominated disk model approaches unity, $-\ell_0$ grows exponentially, accretion is increasingly suppressed, and the time required for a binary to ``turn on'' can begin to exceed the binary's lifetime.

We find, however, that despite their low accretion rates, non-accreting systems are not necessarily electromagnetically dark. 
The truncated CBDs have a suppressed bolometric luminosity and a strongly reduced high-energy, photo-ionizing continuum compared to a standard single-BH \(\alpha\)-disk.
For our $10^6 \Msun$ binary with a few-year orbital period,
the heated inner edge of the CBD can remain luminous at optical and near-infrared frequencies at low-to-moderate redshifts.
As the inner CBD continues to relax and transmit a growing angular momentum current, we expect these optical/NIR luminosities to increase, and the SED peak to shift gradually blueward.
In particular, we find that the non-accreting solutions can be comparable to, or even brighter than, equivalent single-BH $\alpha$-disks in the LSST and the Roman observing bands, while generally retaining sufficient variability for detection as periodic sources. 
We suggest that non-accreting SMBHBs need not be absent from optical/IR surveys, but may instead appear as periodically variable, relatively red systems with weak ionizing continua and an intrinsic lack of X-rays. This provides a possible connection to weak-line quasars as well as potential changing-look phenomena from intermittent minidisk feeding.

We also introduce and validate an ``adiabatic torque-free'' sink prescription that corrects spurious internal-energy losses associated with torque-free mass removal in previous energy-evolving simulations. 
This modification yields stable single-BH disk solutions outside the sink and reduces numerical artifacts that would otherwise contaminate minidisk thermodynamics and inferred emission in binary calculations. 

Several limitations remain. We have considered vertically averaged, un-magnetized, gas-pressure dominated disks with simplified blackbody cooling and electron-scattering opacity. Radiation pressure, MHD turbulence, magnetic pressure support, three-dimensional stream geometry, more realistic opacities, binary eccentricity, and unequal mass ratios may all modify the inner-disk thermodynamics and stream-capture efficiency. The simulations also probe only the early viscous evolution of the non-accreting state, so longer integrations or reduced models will be required to follow the full transition to steady accretion.

Together, these results motivate future work to understand in more detail the extent to which truncated binary accretion and long lived non-accretion phases represent a meaningful mode of binary evolution as it pertains to electromagnetic searches for SMBHBs, the rates and cosmological properties of gravitational wave driven SMBHB mergers, and the future of multi-messenger astronomy with super-massive black holes.

%
% -----------------------------------------------------------------------------
% \vspace{-10pt}
\begin{acknowledgements}
This work was supported by the European Union’s Horizon 2023 research and innovation program under Marie Sklodowska-Curie grant agreement No. 101148364, and
by Sapere Aude Starting grant No. 121587 through the Danish Independent Research Fund. It was additionally supported by NSF AAG No. 2511544.
CT acknowledges many helpful conversations on binary accretion at the workshop \emph{Climbing Cosmic Peaks} through the Sexten Center for Astrophysics Riccardo Giacconi. 
In particular, we thank Zoltan Haiman and Massimo Dotti for discussions on optical time-domain searches; Andrew MacFadyen, Alessia Franchini, Magdalena Siwek, and Mark Avara for conversations on binary accretion dynamics; Alessandra De Rosa for an insightful question on ionizing emission; and Jessie Runnoe for slope-side discussions on AGN broad-line regions.
CT additionally thanks Jonathan Zrake for helpful comments on this manuscript and the anonymous reviewer for an engaging and insightful report.
This work was conceived of, performed, and written by humans: CT, DON, DJD.
Computation time was supported through the Tycho supercomputer hosted at the SCIENCE HPC center at the University of Copenhagen.
% \vspace{10pt}
\end{acknowledgements}

%
% -----------------------------------------------------------------------------
\software{
  \texttt{Sailfish} \citep{Sailfish:2024},
  \texttt{numpy} \citep{numpy},
  \texttt{scipy} \citep{scipy},
  \texttt{matplotlib} \citep{matplotlib},
  \texttt{cmasher} \citep{cmasher}
}
%

%
% ======================================================================
\appendix 
\renewcommand\theequation{A\arabic{equation}}
\renewcommand\thefigure{A\arabic{figure}}
\setcounter{equation}{0}
\setcounter{figure}{0}

%
% ------------------------------------------------------
\vspace{-10pt}
\section{Adiabatic torque-free sinks} \label{app:energy-sink}

The mass and momentum sink terms around each point mass $j$ in our simulations are given, respectively, as
\begin{align}
 \label{eq:sigsink}
    S_{\Sigma, j}     &= -s \Sigma w_j \\
    \mathbf{S_{v, j}} &= -s \Sigma \mathbf{v}_j^\star  w_j
 \label{eq:vsink}
\end{align}
where $s$ has units of inverse time and
$w_j$ is a window function defining the strength of the sink as a function of distance from the sink center.
For torque-free sinks, $\mathbf{v_j^\star} = (\mathbf{v} - \mathbf{v_j}) \cdot \mathbf{\hat r_j} + \mathbf{v_j}$, with $\mathbf{v}$ the gas velocity in the inertial frame, $\mathbf{v_j}$ the component velocity in the same frame, and $\mathbf{\hat r_j}$ the radial unit-vector of the coordinate system centered on binary component $j$.
In this way, the sinks only remove radial momentum in the frame of each component \citep{Dempsey2020, Dittmann+Ryan2021}. 
We take the window function for the single point mass tests to be
\begin{align}
    w_j = \left( 1 - r_j / r_\sink \right)^2 
\end{align}
when $r_j < r_\sink$, and zero otherwise. We adjust this to the smoother $w_j = \exp{\left[ -( r_j / r_\sink )^4 \right]}$, applied to all cells in the vicinity of the sink, for the binary simulations.

\begin{figure*}[t!]
    \centering
    \includegraphics{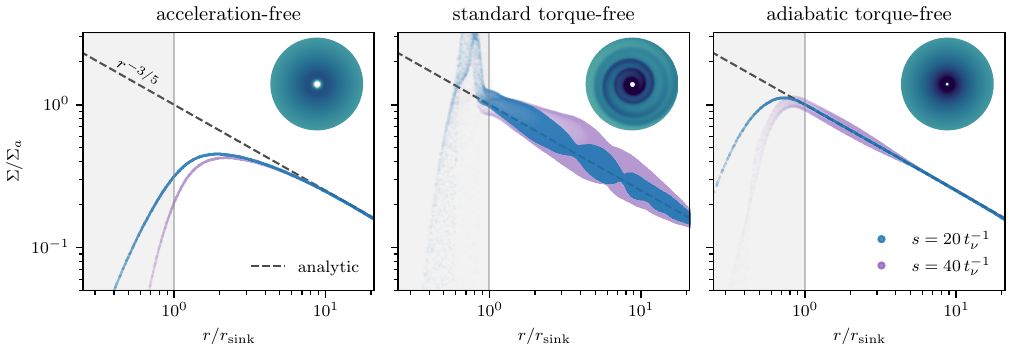}
    \caption{
    Radial surface density profiles in units of $r_\sink$ for simulations around a single point mass for three different sink implementations and two values of the sink rate each.
    Each point in the scatter plot represents a single cell, and vertical spread in the points indicates non-axisymmetry in the solution.
    This can be additionally visualized in the density snapshots given in the top right of each panel.
    The shaded region illustrates the unphysical region interior to $r_\sink$.
    }
    \label{fig:sinks}
\end{figure*}

For the energy sink function, $S_E$, the commonly employed
\begin{align}
    \bar S_{E, j} = -s \Sigma \left(\epsilon + \frac{1}{2} {v_j^\star}^2 \right) \ w_j
 \label{eq:esink_old}
\end{align}
corresponds to spurious removals (or injections) of internal energy and leads to significant numerical artifacts out to tens of $r_s$, demonstrated below.
We instead employ an energy sink of the form 
\begin{align}
    S_{E, j} = -s \Sigma \left(\epsilon + \frac{1}{2} {v_j^\star}^2 - \frac{1}{2} {v_j^\dagger}^2 \right) \ w_j 
 \label{eq:esink_code}
\end{align}
where $\mathbf{v_j^\dagger} = \mathbf{v} - \mathbf{v_j^\star}$. 
To illustrate the difference between the commonly employed term in Eq.~\eqref{eq:esink_old} and that implemented in this work, Eq.~\eqref{eq:esink_code},
consider a coordinate frame centered on a given binary component such that the local fluid velocity in cylindrical coordinates is
$\mathbf{u} = u_r \hat r + u_\varphi \hat \varphi$.
For sinks that remove mass as 
\begin{align}
    \frac{\partial \Sigma}{\partial t} = -s \Sigma
 \label{eq:dsigdt}
\end{align}
the torque-free condition for the removal of linear momentum is
\begin{align}
    \frac{\partial \mathbf{p}}{\partial t} = \mathbf{u} \frac{\partial \Sigma}{\partial t} + \Sigma \frac{\partial \mathbf{u}}{\partial t} = -s \Sigma \mathbf{u^\star}
 \label{eq:dpdt}
\end{align}
where $\mathbf{u^\star} \equiv u_r\ \hat r$ ensures that the sink only removes radial momentum in its own frame.
Therefore, application of the torque-free condition generally corresponds to an increase in the sink-frame rotational velocity
\begin{align}
    \frac{\partial \mathbf{u}}{\partial t} = - s ( \mathbf{u^\star} - \mathbf{u} ) = s\, u_\varphi \hat \varphi \ .
 \label{eq:dudt}
\end{align}
Assuming that the sink does not alter the specific internal energy of the fluid element, the change in energy is then, necessarily,
\begin{align}
    \frac{\partial E}{\partial t} \bigg|_\epsilon &= \epsilon \frac{\partial \Sigma}{\partial t} + \frac{1}{2} \frac{\partial \Sigma}{\partial t} u^2 + \Sigma\, \mathbf{u} \cdot \frac{\partial \mathbf{u}}{\partial t} \\
    &= \epsilon \frac{\partial \Sigma}{\partial t} + \frac{1}{2} \frac{\partial \Sigma}{\partial t} {u^\star}^2 - \frac{1}{2}\frac{\partial \Sigma}{\partial t} u_\varphi^2 \ .
 \label{eq:esink_analytic}
\end{align}
The opposite sign of the final term illustrates how the energy change must include the increased rotational velocity from the torque-free condition.
Therefore, the application of the historically implemented
\begin{align}
    \frac{\partial  E}{\partial t} \rightarrow \epsilon \frac{\partial \Sigma}{\partial t} + \frac{1}{2} \frac{\partial \Sigma}{\partial t} {u^\star}^2 \ ,
 \label{eq:etilde}
\end{align}
does not, in general, correspond with a preservation of $\epsilon$, and rather leads to spontaneous cooling.

To demonstrate this, we perform a set of simulations following the description in Section~\ref{s:single-mass-tests} for three different formulations of $S_E$. 
The first is for an ``acceleration free'' sink where $\mathbf{v^\star_j} \rightarrow \mathbf{v_j}$ in Eqs.~\eqref{eq:vsink}~\&~\eqref{eq:esink_old}.
The second is the ``standard torque-free'' sink given by Eq.~\eqref{eq:esink_old}, and the third is the ``adiabatic torque-free'' sink in Eq.~\eqref{eq:esink_code}.
The surface density for every cell from each of these tests is illustrated as a scatter plot in Figure~\ref{fig:sinks} as a function of radius in units of $r_\sink$, and the inset shows 2D surface density maps of the solutions.
The black dashed curve shows the analytic power-law dependence.
We see that the acceleration free sinks yield smooth axisymmetric solutions (characterized by the lack of spread in the scatter plot), but that these deviate from the analytic expectation out to $\sim 10\, r_s$ due to the torque applied by the sink from the removal of angular momentum.
The spread in the second panel illustrates how the implicit changes to the gas specific internal energy imposed by Eq.~\eqref{eq:esink_old} drive propagating non-axisymmetric waves into the solution out to $\gtrsim 20\, r_s$.
In a binary accretion simulation, this would correspond to the entire circum-single disk, or minidisk, around each binary component, potentially leaving significant imprints in the minidisk morphology, the measured accretion rate and binary torque, and the implied electromagnetic emission.
The final panel, shows our implementation of Eq.~\eqref{eq:esink_code}, which remains axisymmetric (with the exception of some small non-axisymmetry from cat-eye--like features in the fast sink solutions) and accurate for all radii exterior to the sink.

%
% ------------------------------------------------------
\section{Interpretation of the torque parameter, $\ell_0$} 
\label{app:1d-picture}

The vertically and azimuthally averaged disk equations for mass and angular momentum conservation can be written
\begin{align}
    2\pi r \partial_t \Sigma = \partial_r \Mdot \qquad\textrm{and}\qquad
    \Mdot \partial_r \ell = \partial_r F_J - \partial_r T_{\rm int}
 \label{eq:1d-disk}
\end{align}
where $\ell(r) = \sqrt{GMr}$ is the Keplerian specific angular momentum, $F_J$ is the \emph{outward} viscous angular momentum flux due to the $r-\phi$ component of disk stress, and $T_{\rm int}$ is the interaction torque between the disk and the binary potential.
We consider steady-state solutions where, at some inner radius $r_{\rm in}$, the viscous stress disappears, i.e., $F_J(r_{\rm in}) = 0$. 
This can correspond to the inner-edge of a circumbinary disk.
If we assume that the interaction torque is highly localized to the vicinity of $r_{\rm in}$ such that $\int_{r_{\rm in}}^r \partial_r T_{\rm int} \mathrm{d}r \simeq T_{\rm bin}$, where $T_{\rm bin}$ is the total tidal torque exerted on the disk,
integrating Eq.~\eqref{eq:1d-disk} yields
\begin{align}
    \Mdot \left[ \ell(r) - \ell(r_{\rm in}) \right] = F_J(r) - T_{\rm bin} \ .
\end{align}
The viscous angular momentum flux through the disk can then be expressed
\begin{align}
    F_J(r) = \Mdot \ell(r) - \dot J_b
\end{align}
where $\dot J_b = \Mdot \ell(r_{\rm in}) - T_{\rm bin}$ is the net transfer of angular momentum from the disk to the binary (where we've assumed all mass and angular momentum advected across the inner-boundary is deposited onto the binary).
The torque parameter, then, is
\begin{align}
    \ell_0 = \frac{\dot J_b}{\dot M \ell(a)} = \frac{\Mdot \ell(r_{\rm in}) - T_{\rm bin}}{\Mdot a^2 \Omega_b} \ .
 \label{eq:1d-ell0}
\end{align}
One can see that in the limit where the binary exerts no torque (e.g., is only a central potential), $\dot J_b = \Mdot \ell(r_{\rm in})$ and $\ell_0 = \ell(r_{\rm in}) / \ell(a) = \sqrt{r_{\rm in} / a}$.
This corresponds with the typical boundary term from the derivation of \citet{SS1973}.
Thus, the torque parameter coincides with the specific angular momentum of the disk's inner edge when the total torque is dominated by the advection of angular momentum across the inner edge (corresponding to a very efficient transfer of material from the inner-disk into circum-single minidisks).
Of note, this implies an inward angular momentum flux and spin-up (outspiral) of the central object (binary).

However, in the presence of a tidal torque, one can readily see from Eq.~\eqref{eq:1d-ell0} that the torque parameter is not, in general, the specific angular momentum associated to the inner-disk, but instead depends on the magnitude of the integrated interaction torque.
$T_{\rm bin}$ will generally be of order $\Mdot_{\rm int} \ell(r_{\rm in})$, where $\Mdot_{\rm int}$ is the flux of mass that strongly interacts with the binary.
We comment that $\Mdot_{\rm int}$ can be arbitrarily large and does not violate any condition determining the relative location of $r_{\rm in}$ via tidal-viscous torque balance, because both the interaction torque and the viscous torque are linear in the local mass density.
In the non-accreting limit where $\Mdot_{\rm int} \gg \Mdot$, one finds that $\ell_0 \simeq - \Mdot_{\rm int}/\Mdot$ (up to a factor of $\sqrt{r_{\rm in} / a})$ and can immediately appreciate that $(-\ell_0)^{-1}$ sets the efficiency at which the interaction mass is converted into actual accreted mass.
In the 2D picture, then, $-\ell_0$ is a \emph{recycling number}.
It characterizes the average number of times a fluid element must interact strongly with the binary---in say, an accretion stream---before it is actually accreted.

In this context we speculate a distinction between inspiral--outspiral circumbinary accretion solutions: namely, those solutions inducing binary outpsiral are \emph{advective-torque dominated}, as a result of very efficient mass transfer from the inner-disk on to the binary. 
Conversely, solutions promoting binary orbital decay are \emph{tidal-torque dominated}, where inefficient mass transfer causes material to pile-up against the disk's inner edge, increasing the relative contribution of the gravitational torque.
We further speculate that the ``positive minidisk torque'' typically measured in 2D outspiral solutions \citep[e.g.,][]{MML19, SBCodeComp:2024} corresponds to the advected angular momentum at the inner-disk edge. 
In the full 2D solution, this angular momentum flux across the disk's inner-edge is deposited into the binary orbit via gravitational torque (the minidisk torque) in order for material to shed its remaining angular momentum and join a binary component.
We plan to explore this more quantitatively in the future.

%
% ------------------------------------------------------
\section{Time-evolving disk model} \label{app:time-evolution-model}

We follow the general arguments of \citet{Rafikov2013,Rafikov2016} and \citet{Tiede:2025} (based more foundationally on those of \citealt{LyndenBell-Pringle:1974,Pringle1991,SyerClarke1995,Ivanov99}).
The steady-state surface density profile for a thin, Keplerian disk transporting mass at a rate $\Mdot$ that has relaxed to angular momentum current characterized by $\ell_0$ has the radial profile
\begin{align}
    \Sigma(r, \ell_0) = \frac{\Mdot}{3\pi \nu(r)}f(r,\ell_0)\ , \quad f(r, \ell_0) = 1 - \ell_0 \sqrt{\frac{a}{r}}
 \label{eq:therm-balance}
\end{align}
where $a$ is the characteristic scale of the solutions inner-edge where the torque boundary condition is applied.
For localized kinematic viscosity, the dissipation rate per unit area for a Keplerian flow is the usual
\begin{align}
    D(r) = \frac{3GM \Mdot}{8\pi r^3}f(r,\ell_0) \ .
\end{align}
For standard $\alpha$-disks, one can algebraically solve---as in Eqs.~\eqref{eq:sigma_disk}~\&~\eqref{eq:pressure_disk}---for the steady-state disk profiles.
Assuming a self-similar form for $\nu$, it would scale generally as $\nu \propto (f\Mdot)^m r^n$.
We will assume that $a$, $M$, and $\alpha$ remain constant, but we will allow $\Mdot$ to depend on time, provided it changes slowly compared to $t_\nu$. 
This yields surface density and scale height profiles,
\begin{align}
 \label{eq:steady-sigma}
    \Sigma(r, t;\, \ell_0) &= \bar \Sigma \left[f(r, \ell_0)\, \dot M(t) \right]^{1-m} r^{-n} \\
    h(r, t;\, \ell_0) &= \bar h \left[f(r, \ell_0)\, \dot M(t) \right]^{m/2} r^{(2n + 3)/4} \ ,
\label{eq:steady-height}
\end{align}
that vary in time only through $\Mdot$.
$\bar \Sigma$ and $\bar h$ are constants of proportionality set by $\alpha$, $a$, $M$, and fundamental physical constants.
In our case, for a gas-pressure dominated equation of state and electron-scattering opacity, $m = 2/5$ and $n = 3/5$.

Consider now, a two zone accretion flow in which an inner-region has viscously adjusted to angular momentum current $\ell_0 \dot M$, while some outer region has yet to become aware of the inner binary torque.
If the outer region initially carries no net angular momentum current, it has surface density profile $\bar \Sigma_n(r)\, \Mdot_\infty^{1-m}$, where we've absorbed the radial dependence into $\bar \Sigma_n(r) \equiv \bar \Sigma\, r^{-n}$.
The radius where the two zones meet, the radius of influence $r_\nu(t)$, grows in time as the binary torque is communicated outwards by viscous stresses.
The time-dependence of the radius of influence is defined implicitly as $t_{\rm visc}(r_\nu(t)) = t$, where $t_{\rm visc} = 2r^2 / (3\nu)$.
By imposing the time-dependent boundary condition
\begin{align}
    \Sigma\big(r_\nu(t), t;\, \ell_0 \big) = \bar \Sigma_n \big( r_\nu(t) \big)\, \Mdot_\infty^{1-m}
\end{align}
at the radius of influence, and assuming that $\ell_0$ is constant, Eq.~\eqref{eq:steady-sigma} yields the time evolution for the inner-region accretion rate
\begin{align}
    \Mdot(t) = \frac{\Mdot_\infty}{f(r_\nu, \ell_0)} = \frac{\Mdot_\infty}{1 - \ell_0 (t / t_\nu)^{-\lambda}} \ , \quad \lambda = \frac{1}{4 - 2n} \ .
 \label{eq:lambda}
\end{align}
Thus, for some radius in the inner-region $r_{\rm in} < r_\nu$, so long as $\Mdot \ll \Mdot_\infty$, the dissipation rate increases proportional to $\Mdot(t) \sim t^\lambda$, the scale height grows like $h(r_{\rm in}, t) \sim t^{\lambda m / 2}$, and the effective temperature as $T_\eff(r_{\rm in}, t) \sim t^{\lambda / 4}$.

%
% ------------------------------------------------------
\section{Cooling rate comparison} 
\label{app:compression-cooling}
\begin{figure*}[t!]
    \centering
    \includegraphics{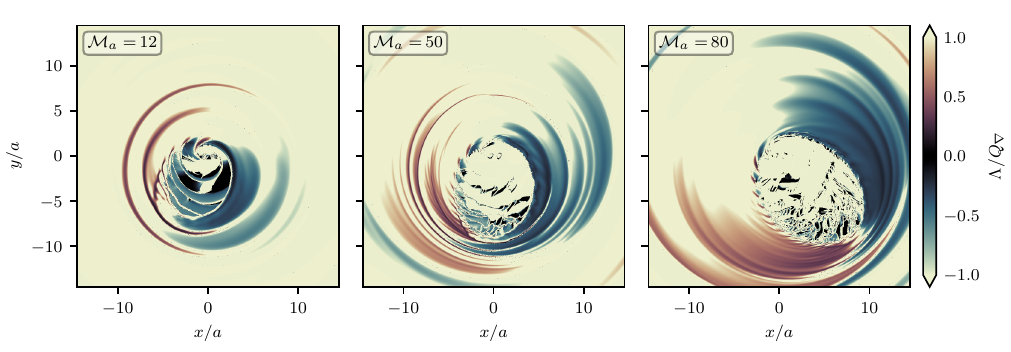}
    \caption{
    Ratios of the blackbody cooling rate with the rate of compressional heating. 
    The cream-colored regions have their energy losses dominated by radiative cooling, while any colored region has its temperature change regulated by converging/diverging fluid orbits.
    }
    \label{fig:compression-v-cooling}
\end{figure*}

The general picture argued here and in \citet{Tiede:2025} is that long-lived binary non-accretion phases occur when accretion streams form from relatively cold gas. 
As a result, they lack sufficient transverse pressure support to resist tidal filamentation and  inefficiently deliver mass to individual binary components.
We observed that when one allows the accretion flow to thermally self-regulate, despite the presence of heated material at the far end of the circumbinary cavity, streams continue to form from relatively cold gas.
In our simulations, the two primary methods for cooling this material orbiting at the cavity edge are radiative blackbody cooling and hydrodynamic expansion.
To determine which regulates the cold stream launching, we plot in Figure~\ref{fig:compression-v-cooling} the ratio of the blackbody cooling rate $\Lambda$ (Eq.~\ref{eq:lambda-cool}) with the ``compressional heating'' rate $Q_\nabla \equiv -P\, \mathbf{\nabla} \cdot \mathbf{v}$, which equates to cooling when negative.
We plot the same snapshots as in Figure~\ref{fig:snapshots} and see that the material that orbits (counter-clockwise) 
from cavity apocenter to pericenter has it's cooling dominated by expansion.
We interpret this as the spreading of nested, eccentric fluid orbits in the inner-disk.
We verify that the same features are present when comparing with the viscous heating rate, and so
it also appears that the hotspot at cavity apocenter may be due to the convergence of orbiting fluid elements in an ``eccentric traffic jam'' \citep{Ragusa2020}.
We plan to characterize in more detail the disk eccentricity profiles in future work.

%
% -----------------------------------------------------------------------------
\bibliographystyle{aasjournalv7}
\bibliography{refs}

@ARTICLE{Liu+2019,
       author = {{Liu}, T. and {Gezari}, S. and {Ayers}, M. and {Burgett}, W. and {Chambers}, K. and {Hodapp}, K. and {Huber}, M.~E. and {Kudritzki}, R.-P. and {Metcalfe}, N. and {Tonry}, J. and {Wainscoat}, R. and {Waters}, C.},
        title = "{Supermassive Black Hole Binary Candidates from the Pan-STARRS1 Medium Deep Survey}",
      journal = {\apj},
         year = 2019,
        month = oct,
       volume = {884},
       number = {1},
          eid = {36},
        pages = {36},
          doi = {10.3847/1538-4357/ab40cb},
archivePrefix = {arXiv},
       eprint = {1906.08315},
 primaryClass = {astro-ph.HE},
       adsurl = {https://ui.adsabs.harvard.edu/abs/2019ApJ...884...36L}
}

@ARTICLE{bogdanovicLRR+2022,
       author = {{Bogdanovi{\'c}}, Tamara and {Miller}, M. Coleman and {Blecha}, Laura},
        title = "{Electromagnetic counterparts to massive black-hole mergers}",
      journal = {Living Reviews in Relativity},
         year = 2022,
        month = dec,
       volume = {25},
       number = {1},
          eid = {3},
        pages = {3},
          doi = {10.1007/s41114-022-00037-8},
archivePrefix = {arXiv},
       eprint = {2109.03262},
 primaryClass = {astro-ph.HE},
       adsurl = {https://ui.adsabs.harvard.edu/abs/2022LRR....25....3B}
}

@ARTICLE{EnnoggiKrolik+2025,
       author = {{Ennoggi}, Lorenzo and {Campanelli}, Manuela and {Zlochower}, Yosef and {Noble}, Scott C. and {Krolik}, Julian and {Cattorini}, Federico and {Kalinani}, Jay V. and {Mewes}, Vassilios and {Chabanov}, Michail and {Ji}, Liwei and {de Simone}, Maria Chiara},
        title = "{Relativistic gas accretion onto supermassive black hole binaries from inspiral through merger}",
      journal = {\prd},
         year = 2025,
        month = sep,
       volume = {112},
       number = {6},
          eid = {063009},
        pages = {063009},
          doi = {10.1103/yc25-v1q4},
archivePrefix = {arXiv},
       eprint = {2502.06389},
 primaryClass = {astro-ph.HE},
       adsurl = {https://ui.adsabs.harvard.edu/abs/2025PhRvD.112f3009E}
}

@ARTICLE{ClyburnZrake:2026,
       author = {{Clyburn}, Madeline and {Zrake}, Jonathan},
        title = "{Unequal Mass Binary Evolution Driven by High Mach Circumbinary Disks}",
      journal = {\mnras},
         year = 2026,
        month = mar,
          doi = {10.1093/mnras/stag567},
       adsurl = {https://ui.adsabs.harvard.edu/abs/2026MNRAS.tmp..552C}
}

@BOOK{ChandrasekharBook61,
       author = {{Chandrasekhar}, Subrahmanyan},
        title = "{Hydrodynamic and hydromagnetic stability}",
         year = 1961,
       adsurl = {https://ui.adsabs.harvard.edu/abs/1961hhs..book.....C}
}

@ARTICLE{Graham+2015,
   author = {{Graham}, M.~J. and {Djorgovski}, S.~G. and {Stern}, D. and 
	{Drake}, A.~J. and {Mahabal}, A.~A. and {Donalek}, C. and {Glikman}, E. and 
	{Larson}, S. and {Christensen}, E.},
    title = "{A systematic search for close supermassive black hole binaries in the Catalina Real-time Transient Survey}",
  journal = {\mnras},
archivePrefix = "arXiv",
   eprint = {1507.07603},
     year = 2015,
    month = oct,
   volume = 453,
    pages = {1562-1576},
      doi = {10.1093/mnras/stv1726},
   adsurl = {http://adsabs.harvard.edu/abs/2015MNRAS.453.1562G}
}

@ARTICLE{BarnesHernquist1996,
   author = {{Barnes}, J.~E. and {Hernquist}, L.},
    title = "{Transformations of Galaxies. II. Gasdynamics in Merging Disk Galaxies}",
  journal = {\apj},
     year = 1996,
    month = nov,
   volume = 471,
    pages = {115},
      doi = {10.1086/177957},
   adsurl = {http://adsabs.harvard.edu/abs/1996ApJ...471..115B}
}

@ARTICLE{Cuadra2009,
   author = {{Cuadra}, J. and {Armitage}, P.~J. and {Alexander}, R.~D. and 
	{Begelman}, M.~C.},
    title = "{Massive black hole binary mergers within subparsec scale gas discs}",
  journal = {\mnras},
archivePrefix = "arXiv",
   eprint = {0809.0311},
     year = 2009,
    month = mar,
   volume = 393,
    pages = {1423-1432},
      doi = {10.1111/j.1365-2966.2008.14147.x},
   adsurl = {http://adsabs.harvard.edu/abs/2009MNRAS.393.1423C}
}

@ARTICLE{Armitage2002,
   author = {{Armitage}, P.~J. and {Natarajan}, P.},
    title = "{Accretion during the Merger of Supermassive Black Holes}",
  journal = {\apjl},
   eprint = {astro-ph/0201318},
     year = 2002,
    month = mar,
   volume = 567,
    pages = {L9-L12},
      doi = {10.1086/339770},
   adsurl = {http://adsabs.harvard.edu/abs/2002ApJ...567L...9A}
}

@ARTICLE{Dorazio2013,
   author = {{D'Orazio}, D.~J. and {Haiman}, Z. and {MacFadyen}, A.},
    title = "{Accretion into the central cavity of a circumbinary disc}",
  journal = {\mnras},
archivePrefix = "arXiv",
   eprint = {1210.0536},
 primaryClass = "astro-ph.GA",
     year = 2013,
    month = dec,
   volume = 436,
    pages = {2997-3020},
      doi = {10.1093/mnras/stt1787},
   adsurl = {http://adsabs.harvard.edu/abs/2013MNRAS.436.2997D}
}

@ARTICLE{MacFadyen2008,
   author = {{MacFadyen}, A.~I. and {Milosavljevi{\'c}}, M.},
    title = "{An Eccentric Circumbinary Accretion Disk and the Detection of Binary Massive Black Holes}",
  journal = {\apj},
   eprint = {astro-ph/0607467},
     year = 2008,
    month = jan,
   volume = 672,
      eid = {83-93},
    pages = {83-93},
      doi = {10.1086/523869},
   adsurl = {http://adsabs.harvard.edu/abs/2008ApJ...672...83M}
}

@ARTICLE{MML17,
   author = {{Miranda}, R. and {Mu{\~n}oz}, D.~J. and {Lai}, D.},
    title = "{Viscous hydrodynamics simulations of circumbinary accretion discs: variability, quasi-steady state and angular momentum transfer}",
  journal = {\mnras},
archivePrefix = "arXiv",
   eprint = {1610.07263},
 primaryClass = "astro-ph.SR",
     year = 2017,
    month = apr,
   volume = 466,
    pages = {1170-1191},
      doi = {10.1093/mnras/stw3189},
   adsurl = {http://adsabs.harvard.edu/abs/2017MNRAS.466.1170M}
}

@ARTICLE{MML19,
   author = {{Mu{\~n}oz}, D.~J. and {Miranda}, R. and {Lai}, D.},
    title = "{Hydrodynamics of Circumbinary Accretion: Angular Momentum Transfer and Binary Orbital Evolution}",
  journal = {\apj},
archivePrefix = "arXiv",
   eprint = {1810.04676},
 primaryClass = "astro-ph.HE",
     year = 2019,
    month = jan,
   volume = 871,
      eid = {84},
    pages = {84},
      doi = {10.3847/1538-4357/aaf867},
   adsurl = {http://adsabs.harvard.edu/abs/2019ApJ...871...84M}
}

@ARTICLE{Noble+2012,
   author = {{Noble}, S.~C. and {Mundim}, B.~C. and {Nakano}, H. and {Krolik}, J.~H. and 
	{Campanelli}, M. and {Zlochower}, Y. and {Yunes}, N.},
    title = "{Circumbinary Magnetohydrodynamic Accretion into Inspiraling Binary Black Holes}",
  journal = {\apj},
archivePrefix = "arXiv",
   eprint = {1204.1073},
 primaryClass = "astro-ph.HE",
     year = 2012,
    month = aug,
   volume = 755,
      eid = {51},
    pages = {51},
      doi = {10.1088/0004-637X/755/1/51},
   adsurl = {http://adsabs.harvard.edu/abs/2012ApJ...755...51N}
}

@ARTICLE{Rafikov2013,
       author = {{Rafikov}, Roman R.},
        title = "{Structure and Evolution of Circumbinary Disks around Supermassive Black Hole Binaries}",
      journal = {\apj},
         year = 2013,
        month = sep,
       volume = {774},
       number = {2},
          eid = {144},
        pages = {144},
          doi = {10.1088/0004-637X/774/2/144},
archivePrefix = {arXiv},
       eprint = {1205.5017},
 primaryClass = {astro-ph.GA},
       adsurl = {https://ui.adsabs.harvard.edu/abs/2013ApJ...774..144R}
}

@ARTICLE{Rafikov2016,
   author = {{Rafikov}, R.~R.},
    title = "{Accretion and Orbital Inspiral in Gas-assisted Supermassive Black Hole Binary Mergers}",
  journal = {\apj},
archivePrefix = "arXiv",
   eprint = {1602.05206},
     year = 2016,
    month = aug,
   volume = 827,
      eid = {111},
    pages = {111},
      doi = {10.3847/0004-637X/827/2/111},
   adsurl = {http://adsabs.harvard.edu/abs/2016ApJ...827..111R}
}

@ARTICLE{SS1973,
   author = {{Shakura}, N.~I. and {Sunyaev}, R.~A.},
    title = "{Black holes in binary systems. Observational appearance.}",
  journal = {\aap},
     year = 1973,
   volume = 24,
    pages = {337-355},
   adsurl = {http://adsabs.harvard.edu/abs/1973A%26A....24..337S}
}

@ARTICLE{SyerClarke1995,
   author = {{Syer}, D. and {Clarke}, C.~J.},
    title = "{Satellites in discs: regulating the accretion luminosity}",
  journal = {\mnras},
   eprint = {astro-ph/9505021},
     year = 1995,
    month = dec,
   volume = 277,
    pages = {758-766},
      doi = {10.1093/mnras/277.3.758},
   adsurl = {http://adsabs.harvard.edu/abs/1995MNRAS.277..758S}
}

@article{Shi+2012,
  author={Ji-Ming Shi and Julian H. Krolik and Stephen H. Lubow and John F. Hawley},
  title={Three-dimensional Magnetohydrodynamic Simulations of Circumbinary Accretion Disks: Disk Structures and Angular Momentum Transport},
  journal={\apj},
  volume={749},
  number={2},
  pages={118},
  url={http://stacks.iop.org/0004-637X/749/i=2/a=118},
  year={2012},
}

@ARTICLE{Shi2015,
       author = {{Shi}, Ji-Ming and {Krolik}, Julian H.},
        title = "{Three-dimensional MHD Simulation of Circumbinary Accretion Disks. II. Net Accretion Rate}",
      journal = {\apj},
         year = 2015,
        month = jul,
       volume = {807},
       number = {2},
          eid = {131},
        pages = {131},
          doi = {10.1088/0004-637X/807/2/131},
archivePrefix = {arXiv},
       eprint = {1503.05561},
 primaryClass = {astro-ph.HE},
       adsurl = {https://ui.adsabs.harvard.edu/abs/2015ApJ...807..131S}
}

@ARTICLE{HKM09,
   author = {{Haiman}, Z. and {Kocsis}, B. and {Menou}, K.},
    title = "{The Population of Viscosity- and Gravitational Wave-driven Supermassive Black Hole Binaries Among Luminous Active Galactic Nuclei}",
  journal = {\apj},
archivePrefix = "arXiv",
   eprint = {0904.1383},
 primaryClass = "astro-ph.CO",
     year = 2009,
    month = aug,
   volume = 700,
    pages = {1952-1969},
      doi = {10.1088/0004-637X/700/2/1952},
   adsurl = {http://adsabs.harvard.edu/abs/2009ApJ...700.1952H}
}

@ARTICLE{Haiman2009:Erratum,
       author = {{Haiman}, Zolt{\'a}n and {Kocsis}, Bence and {Menou}, Kristen},
        title = "{Erratum: ``The Population of Viscosity- and Gravitational Wave-driven Supermassive Black Hole Binaries among Luminous Active Galactic Nuclei'' (2009, ApJ, 700, 1952)}",
      journal = {\apj},
         year = 2022,
        month = oct,
       volume = {937},
       number = {2},
          eid = {129},
        pages = {129},
          doi = {10.3847/1538-4357/ac93f7},
       adsurl = {https://ui.adsabs.harvard.edu/abs/2022ApJ...937..129H}
}

@ARTICLE{Begel:Blan:Rees:1980,
   author = {{Begelman}, M.~C. and {Blandford}, R.~D. and {Rees}, M.~J.},
    title = "{Massive black hole binaries in active galactic nuclei}",
  journal = {\nat},
     year = 1980,
    month = sep,
   volume = 287,
    pages = {307-309},
      doi = {10.1038/287307a0},
}

@article{Ivanov99,
author = {P. B Ivanov and J. C. B Papaloizou and A. G Polnarev}, 
journal = {MNRAS},
title = {The evolution of a supermassive binary caused by an accretion disc},
affiliation = {AA(Queen Mary and Westfield College, University of London, London E1 4NS), AB(Queen Mary and Westfield College, University of London, London E1 4NS), AC(Queen Mary and Westfield College, University of London, London E1 4NS)},
pages = {79},
volume = {307},
year = {1999},
month = {Jul},
doi = {10.1046/j.1365-8711.1999.02623.x},
pmid = {1999MNRAS.307...79I},
uri = {papers://77183D9B-B140-4DA6-B598-A9837E5CF53B/Paper/p2339}
}

@ARTICLE{Volonteri+2003,
   author = {{Volonteri}, M. and {Haardt}, F. and {Madau}, P.},
    title = "{The Assembly and Merging History of Supermassive Black Holes in Hierarchical Models of Galaxy Formation}",
  journal = {\apj},
   eprint = {arXiv:astro-ph/0207276},
     year = 2003,
    month = jan,
   volume = 582,
    pages = {559-573},
      doi = {10.1086/344675},
   adsurl = {http://adsabs.harvard.edu/abs/2003ApJ...582..559V}
}

@ARTICLE{al94,
   author = {{Artymowicz}, P. and {Lubow}, S.~H.},
    title = "{Dynamics of binary-disk interaction. 1: Resonances and disk gap sizes}",
  journal = {\apj},
     year = 1994,
    month = feb,
   volume = 421,
    pages = {651-667},
      doi = {10.1086/173679},
   adsurl = {http://adsabs.harvard.edu/abs/1994ApJ...421..651A}
}

@ARTICLE{Kocsis+2012a,
   author = {{Kocsis}, B. and {Haiman}, Z. and {Loeb}, A.},
    title = "{Gas pile-up, gap overflow and Type 1.5 migration in circumbinary discs: general theory}",
  journal = {\mnras},
archivePrefix = "arXiv",
 primaryClass = "astro-ph.EP",
     year = 2012,
    month = dec,
   volume = 427,
    pages = {2660-2679},
      doi = {10.1111/j.1365-2966.2012.22129.x},
}

@ARTICLE{Kocsis+2012b,
   author = {{Kocsis}, B. and {Haiman}, Z. and {Loeb}, A.},
    title = "{Gas pile-up, gap overflow and Type 1.5 migration in circumbinary discs: application to supermassive black hole binaries}",
  journal = {\mnras},
archivePrefix = "arXiv",
 primaryClass = "astro-ph.HE",
     year = 2012,
    month = dec,
   volume = 427,
    pages = {2680-2700},
      doi = {10.1111/j.1365-2966.2012.22118.x},
}

@ARTICLE{Ragusa+2016,
   author = {{Ragusa}, E. and {Lodato}, G. and {Price}, D.~J.},
    title = "{Suppression of the accretion rate in thin discs around binary black holes}",
  journal = {\mnras},
archivePrefix = "arXiv",
   eprint = {1605.01730},
 primaryClass = "astro-ph.HE",
     year = 2016,
    month = aug,
   volume = 460,
    pages = {1243-1253},
      doi = {10.1093/mnras/stw1081},
   adsurl = {http://adsabs.harvard.edu/abs/2016MNRAS.460.1243R}
}

@ARTICLE{Pringle1991,
   author = {{Pringle}, J.~E.},
    title = "{The properties of external accretion discs}",
  journal = {\mnras},
     year = 1991,
    month = feb,
   volume = 248,
    pages = {754-759},
      doi = {10.1093/mnras/248.4.754},
   adsurl = {http://adsabs.harvard.edu/abs/1991MNRAS.248..754P}
}

@ARTICLE{RyanMacFadyen2017,
   author = {{Ryan}, G. and {MacFadyen}, A.},
    title = "{Minidisks in Binary Black Hole Accretion}",
  journal = {\apj},
archivePrefix = "arXiv",
   eprint = {1611.00341},
 primaryClass = "astro-ph.HE",
     year = 2017,
    month = feb,
   volume = 835,
      eid = {199},
    pages = {199},
      doi = {10.3847/1538-4357/835/2/199},
   adsurl = {http://adsabs.harvard.edu/abs/2017ApJ...835..199R}
}

@article{Farris15,
author = {Farris, Brian D. and Duffell, Paul and MacFadyen, Andrew I. and Haiman, Zoltán},
title = {Binary black hole accretion during inspiral and merger},
journal = {Monthly Notices of the Royal Astronomical Society: Letters},
volume = {447},
number = {1},
pages = {L80},
year = {2015},
doi = {10.1093/mnrasl/slu184},
URL = { + http://dx.doi.org/10.1093/mnrasl/slu184},
eprint = {/oup/backfile/content_public/journal/mnrasl/447/1/10.1093/mnrasl/slu184/2/slu184.pdf}
}

@article{Peters1964,
  title={Gravitational radiation and the motion of two point masses},
  author={Peters, Philip Carl},
  journal={Physical Review},
  volume={136},
  number={4B},
  pages={B1224},
  year={1964},
  publisher={APS}
}

@ARTICLE{Dorazio2016,
       author = {{D'Orazio}, Daniel J. and {Haiman}, Zolt{\'a}n and {Duffell}, Paul and
        {MacFadyen}, Andrew and {Farris}, Brian},
        title = "{A transition in circumbinary accretion discs at a binary mass ratio of 1:25}",
      journal = {\mnras},
         year = 2016,
        month = Jul,
       volume = {459},
        pages = {2379-2393},
          doi = {10.1093/mnras/stw792},
archivePrefix = {arXiv},
       eprint = {1512.05788},
 primaryClass = {astro-ph.HE},
       adsurl = {https://ui.adsabs.harvard.edu/\#abs/2016MNRAS.459.2379D}
}

@ARTICLE{Komossa:review:2006,
       author = {{Komossa}, S.},
        title = "{Observational evidence for binary black holes and active double nuclei}",
      journal = {\memsai},
         year = 2006,
        month = jan,
       volume = {77},
        pages = {733},
       adsurl = {https://ui.adsabs.harvard.edu/abs/2006MmSAI..77..733K}
}

@software{Sailfish:2024,
       author = {{Zrake}, Jonathan and {MacFadyen}, Andrew},
        title = {Sailfish: GPU-accelerated grid-based astrophysics gas dynamics code},
 howpublished = {Astrophysics Source Code Library, record ascl:2408.004},
         year = 2024,
        month = aug,
          eid = {ascl:2408.004},
       adsurl = {https://ui.adsabs.harvard.edu/abs/2024ascl.soft08004Z}
}

@article{Zrake2020,
   title={Equilibrium Eccentricity of Accreting Binaries},
   volume={909},
   ISSN={2041-8213},
   url={http://dx.doi.org/10.3847/2041-8213/abdd1c},
   DOI={10.3847/2041-8213/abdd1c},
   number={1},
   journal={The Astrophysical Journal Letters},
   publisher={American Astronomical Society},
   author={Zrake, Jonathan and Tiede, Christopher and MacFadyen, Andrew and Haiman, Zoltán},
   year={2021},
   month={Mar},
   pages={L13}
}

@ARTICLE{Zrake:CLIs:2025,
       author = {{Zrake}, Jonathan and {Clyburn}, Madeline and {Feyan}, Samuel},
        title = "{Changing-look inspirals: Trends and switches in AGN disk emission as signposts for merging black hole binaries}",
      journal = {\mnras},
         year = 2025,
        month = jan,
          doi = {10.1093/mnras/staf171},
archivePrefix = {arXiv},
       eprint = {2410.04961},
 primaryClass = {astro-ph.HE},
       adsurl = {https://ui.adsabs.harvard.edu/abs/2025MNRAS.tmp..161Z}
}

@ARTICLE{Tiede2020,
       author = {{Tiede}, Christopher and {Zrake}, Jonathan and {MacFadyen}, Andrew and {Haiman}, Zoltan},
        title = "{Gas-driven Inspiral of Binaries in Thin Accretion Disks}",
      journal = {\apj},
         year = 2020,
        month = sep,
       volume = {900},
       number = {1},
          eid = {43},
        pages = {43},
          doi = {10.3847/1538-4357/aba432},
archivePrefix = {arXiv},
       eprint = {2005.09555},
 primaryClass = {astro-ph.GA},
       adsurl = {https://ui.adsabs.harvard.edu/abs/2020ApJ...900...43T}
}

@ARTICLE{Tiede2022,
       author = {{Tiede}, Christopher and {Zrake}, Jonathan and {MacFadyen}, Andrew and {Haiman}, Zolt{\'a}n},
        title = "{How Binaries Accrete: Hydrodynamic Simulations with Passive Tracer Particles}",
      journal = {\apj},
         year = 2022,
        month = jun,
       volume = {932},
       number = {1},
          eid = {24},
        pages = {24},
          doi = {10.3847/1538-4357/ac6c2b},
archivePrefix = {arXiv},
       eprint = {2111.04721},
 primaryClass = {astro-ph.GA},
       adsurl = {https://ui.adsabs.harvard.edu/abs/2022ApJ...932...24T}
}

@ARTICLE{Munoz2020,
       author = {{Mu{\~n}oz}, Diego J. and {Lithwick}, Yoram},
        title = "{Long-lived Eccentric Modes in Circumbinary Disks}",
      journal = {\apj},
         year = 2020,
        month = dec,
       volume = {905},
       number = {2},
          eid = {106},
        pages = {106},
          doi = {10.3847/1538-4357/abc74c},
archivePrefix = {arXiv},
       eprint = {2008.08085},
 primaryClass = {astro-ph.HE},
       adsurl = {https://ui.adsabs.harvard.edu/abs/2020ApJ...905..106M}
}

@ARTICLE{Ragusa2020,
       author = {{Ragusa}, Enrico and {Alexander}, Richard and {Calcino}, Josh and {Hirsh}, Kieran and {Price}, Daniel J.},
        title = "{The evolution of large cavities and disc eccentricity in circumbinary discs}",
      journal = {\mnras},
         year = 2020,
        month = dec,
       volume = {499},
       number = {3},
        pages = {3362-3380},
          doi = {10.1093/mnras/staa2954},
archivePrefix = {arXiv},
       eprint = {2009.10738},
 primaryClass = {astro-ph.EP},
       adsurl = {https://ui.adsabs.harvard.edu/abs/2020MNRAS.499.3362R}
}

@ARTICLE{dAscoliNoble:2018,
       author = {{d'Ascoli}, St{\'e}phane and {Noble}, Scott C. and {Bowen}, Dennis B. and {Campanelli}, Manuela and {Krolik}, Julian H. and {Mewes}, Vassilios},
        title = "{Electromagnetic Emission from Supermassive Binary Black Holes Approaching Merger}",
      journal = {\apj},
         year = 2018,
        month = oct,
       volume = {865},
       number = {2},
          eid = {140},
        pages = {140},
          doi = {10.3847/1538-4357/aad8b4},
archivePrefix = {arXiv},
       eprint = {1806.05697},
 primaryClass = {astro-ph.HE},
       adsurl = {https://ui.adsabs.harvard.edu/abs/2018ApJ...865..140D}
}

@ARTICLE{Noble2021,
       author = {{Noble}, Scott C. and {Krolik}, Julian H. and {Campanelli}, Manuela and {Zlochower}, Yosef and {Mundim}, Bruno C. and {Nakano}, Hiroyuki and {Zilh{\~a}o}, Miguel},
        title = "{Mass-ratio and Magnetic Flux Dependence of Modulated Accretion from Circumbinary Disks}",
      journal = {\apj},
         year = 2021,
        month = dec,
       volume = {922},
       number = {2},
          eid = {175},
        pages = {175},
          doi = {10.3847/1538-4357/ac2229},
archivePrefix = {arXiv},
       eprint = {2103.12100},
 primaryClass = {astro-ph.HE},
       adsurl = {https://ui.adsabs.harvard.edu/abs/2021ApJ...922..175N}
}

@ARTICLE{Avara:2024,
       author = {{Avara}, Mark J. and {Krolik}, Julian H. and {Campanelli}, Manuela and {Noble}, Scott C. and {Bowen}, Dennis and {Ryu}, Taeho},
        title = "{Accretion onto a Supermassive Black Hole Binary before Merger}",
      journal = {\apj},
         year = 2024,
        month = oct,
       volume = {974},
       number = {2},
          eid = {242},
        pages = {242},
          doi = {10.3847/1538-4357/ad5bda},
archivePrefix = {arXiv},
       eprint = {2305.18538},
 primaryClass = {astro-ph.HE},
       adsurl = {https://ui.adsabs.harvard.edu/abs/2024ApJ...974..242A}
}

@ARTICLE{Heath+Nixon2020,
       author = {{Heath}, R.~M. and {Nixon}, C.~J.},
        title = "{On the orbital evolution of binaries with circumbinary discs}",
      journal = {\aap},
         year = 2020,
        month = sep,
       volume = {641},
          eid = {A64},
        pages = {A64},
          doi = {10.1051/0004-6361/202038548},
archivePrefix = {arXiv},
       eprint = {2007.11592},
 primaryClass = {astro-ph.HE},
       adsurl = {https://ui.adsabs.harvard.edu/abs/2020A&A...641A..64H}
}

@ARTICLE{Franchini2021,
       author = {{Franchini}, Alessia and {Sesana}, Alberto and {Dotti}, Massimo},
        title = "{Circumbinary disc self-gravity governing supermassive black hole binary mergers}",
      journal = {\mnras},
         year = 2021,
        month = oct,
       volume = {507},
       number = {1},
        pages = {1458-1467},
          doi = {10.1093/mnras/stab2234},
archivePrefix = {arXiv},
       eprint = {2106.13253},
 primaryClass = {astro-ph.HE},
       adsurl = {https://ui.adsabs.harvard.edu/abs/2021MNRAS.507.1458F}
}

@ARTICLE{Franchini2022,
       author = {{Franchini}, Alessia and {Lupi}, Alessandro and {Sesana}, Alberto},
        title = "{Resolving Massive Black Hole Binary Evolution via Adaptive Particle Splitting}",
      journal = {\apjl},
         year = 2022,
        month = apr,
       volume = {929},
       number = {1},
          eid = {L13},
        pages = {L13},
          doi = {10.3847/2041-8213/ac63a2},
archivePrefix = {arXiv},
       eprint = {2201.05619},
 primaryClass = {astro-ph.HE},
       adsurl = {https://ui.adsabs.harvard.edu/abs/2022ApJ...929L..13F}
}

@ARTICLE{Duffell2020,
       author = {{Duffell}, Paul C. and {D'Orazio}, Daniel and {Derdzinski}, Andrea and {Haiman}, Zoltan and {MacFadyen}, Andrew and {Rosen}, Anna L. and {Zrake}, Jonathan},
        title = "{Circumbinary Disks: Accretion and Torque as a Function of Mass Ratio and Disk Viscosity}",
      journal = {\apj},
         year = 2020,
        month = sep,
       volume = {901},
       number = {1},
          eid = {25},
        pages = {25},
          doi = {10.3847/1538-4357/abab95},
archivePrefix = {arXiv},
       eprint = {1911.05506},
 primaryClass = {astro-ph.SR},
       adsurl = {https://ui.adsabs.harvard.edu/abs/2020ApJ...901...25D}
}

@ARTICLE{Dorazio2021,
       author = {{D'Orazio}, Daniel J. and {Duffell}, Paul C.},
        title = "{Orbital Evolution of Equal-mass Eccentric Binaries due to a Gas Disk: Eccentric Inspirals and Circular Outspirals}",
      journal = {\apjl},
         year = 2021,
        month = jun,
       volume = {914},
       number = {1},
          eid = {L21},
        pages = {L21},
          doi = {10.3847/2041-8213/ac0621},
archivePrefix = {arXiv},
       eprint = {2103.09251},
 primaryClass = {astro-ph.HE},
       adsurl = {https://ui.adsabs.harvard.edu/abs/2021ApJ...914L..21D}
}

@ARTICLE{Dittmann+Ryan2021,
       author = {{Dittmann}, Alexander J. and {Ryan}, Geoffrey},
        title = "{Preventing Anomalous Torques in Circumbinary Accretion Simulations}",
      journal = {\apj},
         year = 2021,
        month = nov,
       volume = {921},
       number = {1},
          eid = {71},
        pages = {71},
          doi = {10.3847/1538-4357/ac1bbd},
archivePrefix = {arXiv},
       eprint = {2102.05684},
 primaryClass = {astro-ph.HE},
       adsurl = {https://ui.adsabs.harvard.edu/abs/2021ApJ...921...71D}
}

@ARTICLE{Dittmann2022,
       author = {{Dittmann}, Alexander J. and {Ryan}, Geoffrey},
        title = "{A survey of disc thickness and viscosity in circumbinary accretion: Binary evolution, variability, and disc morphology}",
      journal = {\mnras},
         year = 2022,
        month = jul,
       volume = {513},
       number = {4},
        pages = {6158-6176},
          doi = {10.1093/mnras/stac935},
archivePrefix = {arXiv},
       eprint = {2201.07816},
 primaryClass = {astro-ph.HE},
       adsurl = {https://ui.adsabs.harvard.edu/abs/2022MNRAS.513.6158D}
}

@ARTICLE{Dempsey2020,
       author = {{Dempsey}, Adam M. and {Mu{\~n}oz}, Diego and {Lithwick}, Yoram},
        title = "{Inner Boundary Condition in Quasi-Lagrangian Simulations of Accretion Disks}",
      journal = {\apjl},
         year = 2020,
        month = apr,
       volume = {892},
       number = {2},
          eid = {L29},
        pages = {L29},
          doi = {10.3847/2041-8213/ab800e},
archivePrefix = {arXiv},
       eprint = {2002.05164},
 primaryClass = {astro-ph.EP},
       adsurl = {https://ui.adsabs.harvard.edu/abs/2020ApJ...892L..29D}
}

@BOOK{RybickiLightman:RPA,
       author = {{Rybicki}, George B. and {Lightman}, Alan P.},
        title = "{Radiative Processes in Astrophysics}",
         year = 1986,
       adsurl = {https://ui.adsabs.harvard.edu/abs/1986rpa..book.....R}
}

@ARTICLE{Penzlin2022,
       author = {{Penzlin}, Anna B.~T. and {Kley}, Wilhelm and {Audiffren}, Hugo and {Sch{\"a}fer}, Christoph M.},
        title = "{Binary orbital evolution driven by a circumbinary disc}",
      journal = {\aap},
         year = 2022,
        month = apr,
       volume = {660},
          eid = {A101},
        pages = {A101},
          doi = {10.1051/0004-6361/202141399},
archivePrefix = {arXiv},
       eprint = {2202.06681},
 primaryClass = {astro-ph.SR},
       adsurl = {https://ui.adsabs.harvard.edu/abs/2022A&A...660A.101P}
}

@ARTICLE{Siwek2023,
       author = {{Siwek}, Magdalena and {Weinberger}, Rainer and {Hernquist}, Lars},
        title = "{Orbital evolution of binaries in circumbinary disks}",
      journal = {\mnras},
         year = 2023,
        month = apr,
          doi = {10.1093/mnras/stad1131},
archivePrefix = {arXiv},
       eprint = {2302.01785},
 primaryClass = {astro-ph.HE},
       adsurl = {https://ui.adsabs.harvard.edu/abs/2023MNRAS.tmp.1097S}
}

@ARTICLE{Valli:2024,
       author = {{Valli}, Ruggero and {Tiede}, Christopher and {Vigna-G{\'o}mez}, Alejandro and {Cuadra}, Jorge and {Siwek}, Magdalena and {Ma}, Jing-Ze and {D'Orazio}, Daniel J. and {Zrake}, Jonathan and {de Mink}, Selma E.},
        title = "{Long-term evolution of binary orbits induced by circumbinary disks}",
      journal = {\aap},
         year = 2024,
        month = aug,
       volume = {688},
          eid = {A128},
        pages = {A128},
          doi = {10.1051/0004-6361/202449421},
archivePrefix = {arXiv},
       eprint = {2401.17355},
 primaryClass = {astro-ph.HE},
       adsurl = {https://ui.adsabs.harvard.edu/abs/2024A&A...688A.128V}
}

@ARTICLE{Westernacher-Schneider:2022,
       author = {{Westernacher-Schneider}, John Ryan and {Zrake}, Jonathan and {MacFadyen}, Andrew and {Haiman}, Zolt{\'a}n},
        title = "{Multiband light curves from eccentric accreting supermassive black hole binaries}",
      journal = {\prd},
         year = 2022,
        month = nov,
       volume = {106},
       number = {10},
          eid = {103010},
        pages = {103010},
          doi = {10.1103/PhysRevD.106.103010},
archivePrefix = {arXiv},
       eprint = {2111.06882},
 primaryClass = {astro-ph.HE},
       adsurl = {https://ui.adsabs.harvard.edu/abs/2022PhRvD.106j3010W}
}

@ARTICLE{Westernacher-Schneider:2023,
       author = {{Westernacher-Schneider}, John Ryan and {Zrake}, Jonathan and {MacFadyen}, Andrew and {Haiman}, Zolt{\'a}n},
        title = "{Characteristic signatures of accreting binary black holes produced by eccentric minidisks}",
      journal = {arXiv e-prints},
         year = 2023,
        month = jul,
          eid = {arXiv:2307.01154},
        pages = {arXiv:2307.01154},
          doi = {10.48550/arXiv.2307.01154},
archivePrefix = {arXiv},
       eprint = {2307.01154},
 primaryClass = {astro-ph.HE},
       adsurl = {https://ui.adsabs.harvard.edu/abs/2023arXiv230701154W}
}

@ARTICLE{Krauth:2023,
       author = {{Krauth}, Luke Major and {Davelaar}, Jordy and {Haiman}, Zolt{\'a}n and {Westernacher-Schneider}, John Ryan and {Zrake}, Jonathan and {MacFadyen}, Andrew},
        title = "{Disappearing thermal X-ray emission as a tell-tale signature of merging massive black hole binaries}",
      journal = {\mnras},
         year = 2023,
        month = dec,
       volume = {526},
       number = {4},
        pages = {5441-5454},
          doi = {10.1093/mnras/stad3095},
archivePrefix = {arXiv},
       eprint = {2304.02575},
 primaryClass = {astro-ph.HE},
       adsurl = {https://ui.adsabs.harvard.edu/abs/2023MNRAS.526.5441K}
}

@ARTICLE{SBCodeComp:2024,
       author = {{Duffell}, Paul C. and {Dittmann}, Alexander J. and {D'Orazio}, Daniel J. and {Franchini}, Alessia and {Kratter}, Kaitlin M. and {Penzlin}, Anna B.~T. and {Ragusa}, Enrico and {Siwek}, Magdalena and {Tiede}, Christopher and {Wang}, Haiyang and {Zrake}, Jonathan and {Dempsey}, Adam M. and {Haiman}, Zoltan and {Lupi}, Alessandro and {Pirog}, Michal and {Ryan}, Geoffrey},
        title = "{The Santa Barbara Binary‑disk Code Comparison}",
      journal = {\apj},
         year = 2024,
        month = aug,
       volume = {970},
       number = {2},
          eid = {156},
        pages = {156},
          doi = {10.3847/1538-4357/ad5a7e},
archivePrefix = {arXiv},
       eprint = {2402.13039},
 primaryClass = {astro-ph.SR},
       adsurl = {https://ui.adsabs.harvard.edu/abs/2024ApJ...970..156D}
}

@ARTICLE{Cocchiararo:2024,
       author = {{Cocchiararo}, F. and {Franchini}, A. and {Lupi}, A. and {Sesana}, A.},
        title = "{Electromagnetic signatures from accreting massive black hole binaries in time domain photometric surveys}",
      journal = {\aap},
         year = 2024,
        month = nov,
       volume = {691},
          eid = {A250},
        pages = {A250},
          doi = {10.1051/0004-6361/202449598},
archivePrefix = {arXiv},
       eprint = {2402.05175},
 primaryClass = {astro-ph.HE},
       adsurl = {https://ui.adsabs.harvard.edu/abs/2024A&A...691A.250C}
}

@ARTICLE{Cocchiararo:2025,
       author = {{Cocchiararo}, Fabiola and {Franchini}, Alessia and {Lupi}, Alessandro and {Sesana}, Alberto},
        title = "{Radiation pressure role in accreting massive black hole binaries}",
      journal = {arXiv e-prints},
         year = 2025,
        month = aug,
          eid = {arXiv:2508.18349},
        pages = {arXiv:2508.18349},
          doi = {10.48550/arXiv.2508.18349},
archivePrefix = {arXiv},
       eprint = {2508.18349},
 primaryClass = {astro-ph.HE},
       adsurl = {https://ui.adsabs.harvard.edu/abs/2025arXiv250818349C}
}

@ARTICLE{WangBaiLai:2023,
       author = {{Wang}, Hai-Yang and {Bai}, Xue-Ning and {Lai}, Dong},
        title = "{On the Role of Dynamical Cooling in the Dynamics of Circumbinary Disks}",
      journal = {\apj},
         year = 2023,
        month = feb,
       volume = {943},
       number = {2},
          eid = {175},
        pages = {175},
          doi = {10.3847/1538-4357/acac77},
archivePrefix = {arXiv},
       eprint = {2212.04199},
 primaryClass = {astro-ph.HE},
       adsurl = {https://ui.adsabs.harvard.edu/abs/2023ApJ...943..175W}
}

@ARTICLE{Most:MHD:2024,
       author = {{Most}, Elias R. and {Wang}, Hai-Yang},
        title = "{Magnetically Arrested Circumbinary Accretion Flows}",
      journal = {\apjl},
         year = 2024,
        month = sep,
       volume = {973},
       number = {1},
          eid = {L19},
        pages = {L19},
          doi = {10.3847/2041-8213/ad7713},
archivePrefix = {arXiv},
       eprint = {2408.00757},
 primaryClass = {astro-ph.HE},
       adsurl = {https://ui.adsabs.harvard.edu/abs/2024ApJ...973L..19M}
}

@ARTICLE{Wang:BMAD:2025,
       author = {{Wang}, Hai-Yang and {Most}, Elias R. and {Hopkins}, Philip F.},
        title = "{$\textit{BMAD}$-Circumbinary Magnetically Arrested Disks around Stellar or Black Hole Binaries: Hot Accretion Flows, Disk Properties, and Angular Momentum Transfer}",
      journal = {arXiv e-prints},
         year = 2025,
        month = aug,
          eid = {arXiv:2508.16855},
        pages = {arXiv:2508.16855},
          doi = {10.48550/arXiv.2508.16855},
archivePrefix = {arXiv},
       eprint = {2508.16855},
 primaryClass = {astro-ph.HE},
       adsurl = {https://ui.adsabs.harvard.edu/abs/2025arXiv250816855W}
}

@ARTICLE{Hopkins:MAD-Quasars:2024,
       author = {{Hopkins}, Philip F. and {Squire}, Jonathan and {Su}, Kung-Yi and {Steinwandel}, Ulrich P. and {Kremer}, Kyle and {Shi}, Yanlong and {Grudic}, Michael Y. and {Wellons}, Sarah and {Faucher-Giguere}, Claude-Andre and {Angles-Alcazar}, Daniel and {Murray}, Norman and {Quataert}, Eliot},
        title = "{FORGE'd in FIRE II: The Formation of Magnetically-Dominated Quasar Accretion Disks from Cosmological Initial Conditions}",
      journal = {The Open Journal of Astrophysics},
         year = 2024,
        month = mar,
       volume = {7},
          eid = {19},
        pages = {19},
          doi = {10.21105/astro.2310.04506},
archivePrefix = {arXiv},
       eprint = {2310.04506},
 primaryClass = {astro-ph.HE},
       adsurl = {https://ui.adsabs.harvard.edu/abs/2024OJAp....7E..19H}
}

@ARTICLE{ChenXin:2020,
       author = {{Chen}, Yu-Ching and {Liu}, Xin and {Liao}, Wei-Ting and {Holgado}, A. Miguel and {Guo}, Hengxiao and {Gruendl}, Robert A. and {Morganson}, Eric and {Shen}, Yue and {Zhang}, Kaiwen and {Abbott}, Tim M.~C. and {Aguena}, Michel and {Allam}, Sahar and {Avila}, Santiago and {Bertin}, Emmanuel and {Bhargava}, Sunayana and {Brooks}, David and {Burke}, David L. and {Carnero Rosell}, Aurelio and {Carollo}, Daniela and {Carrasco Kind}, Matias and {Carretero}, Jorge and {Costanzi}, Matteo and {da Costa}, Luiz N. and {Davis}, Tamara M. and {De Vicente}, Juan and {Desai}, Shantanu and {Diehl}, H. Thomas and {Doel}, Peter and {Everett}, Spencer and {Flaugher}, Brenna and {Friedel}, Douglas and {Frieman}, Joshua and {Garc{\'\i}a-Bellido}, Juan and {Gaztanaga}, Enrique and {Glazebrook}, Karl and {Gruen}, Daniel and {Gutierrez}, Gaston and {Hinton}, Samuel R. and {Hollowood}, Devon L. and {James}, David J. and {Kim}, Alex G. and {Kuehn}, Kyler and {Kuropatkin}, Nikolay and {Lewis}, Geraint F. and {Lidman}, Christopher and {Lima}, Marcos and {Maia}, Marcio A.~G. and {March}, Marisa and {Marshall}, Jennifer L. and {Menanteau}, Felipe and {Miquel}, Ramon and {Palmese}, Antonella and {Paz-Chinch{\'o}n}, Francisco and {Plazas}, Andr{\'e}s A. and {Sanchez}, Eusebio and {Schubnell}, Michael and {Serrano}, Santiago and {Sevilla-Noarbe}, Ignacio and {Smith}, Mathew and {Suchyta}, Eric and {Swanson}, Molly E.~C. and {Tarle}, Gregory and {Tucker}, Brad E. and {Norbert Varga}, Tamas and {Walker}, Alistair R.},
        title = "{Candidate periodically variable quasars from the Dark Energy Survey and the Sloan Digital Sky Survey}",
      journal = {\mnras},
         year = 2020,
        month = dec,
       volume = {499},
       number = {2},
        pages = {2245-2264},
          doi = {10.1093/mnras/staa2957},
archivePrefix = {arXiv},
       eprint = {2008.12329},
 primaryClass = {astro-ph.HE},
       adsurl = {https://ui.adsabs.harvard.edu/abs/2020MNRAS.499.2245C}
}

@ARTICLE{Charisi:2016,
       author = {{Charisi}, M. and {Bartos}, I. and {Haiman}, Z. and {Price-Whelan}, A.~M. and {Graham}, M.~J. and {Bellm}, E.~C. and {Laher}, R.~R. and {M{\'a}rka}, S.},
        title = "{A population of short-period variable quasars from PTF as supermassive black hole binary candidates}",
      journal = {\mnras},
         year = 2016,
        month = dec,
       volume = {463},
       number = {2},
        pages = {2145-2171},
          doi = {10.1093/mnras/stw1838},
archivePrefix = {arXiv},
       eprint = {1604.01020},
 primaryClass = {astro-ph.GA},
       adsurl = {https://ui.adsabs.harvard.edu/abs/2016MNRAS.463.2145C}
}

@ARTICLE{Gaskell:1985,
       author = {{Gaskell}, C.~M.},
        title = "{Galactic mergers, starburst galaxies, quasar activity and massive binary black holes}",
      journal = {\nat},
         year = 1985,
        month = may,
       volume = {315},
       number = {6018},
        pages = {386},
          doi = {10.1038/315386a0},
       adsurl = {https://ui.adsabs.harvard.edu/abs/1985Natur.315..386G}
}

@ARTICLE{Dittmann:q-mach:2024,
       author = {{Dittmann}, Alexander J. and {Ryan}, Geoffrey},
        title = "{The Evolution of Accreting Binaries: From Brown Dwarfs to Supermassive Black Holes}",
      journal = {\apj},
         year = 2024,
        month = may,
       volume = {967},
       number = {1},
          eid = {12},
        pages = {12},
          doi = {10.3847/1538-4357/ad2f1e},
archivePrefix = {arXiv},
       eprint = {2310.07758},
 primaryClass = {astro-ph.GA},
       adsurl = {https://ui.adsabs.harvard.edu/abs/2024ApJ...967...12D}
}

@ARTICLE{Dittmann:2026,
       author = {{Dittmann}, Alexander J. and {Ryan}, Geoffrey and {Combi}, Luciano},
        title = "{Eccentric Binaries Accreting from Thin Disks: Orbital Evolution}",
      journal = {arXiv e-prints},
         year = 2025,
        month = dec,
          eid = {arXiv:2512.11954},
        pages = {arXiv:2512.11954},
          doi = {10.48550/arXiv.2512.11954},
archivePrefix = {arXiv},
       eprint = {2512.11954},
 primaryClass = {astro-ph.GA},
       adsurl = {https://ui.adsabs.harvard.edu/abs/2025arXiv251211954D}
}

@ARTICLE{Tiede:2025,
       author = {{Tiede}, Christopher and {Zrake}, Jonathan and {MacFadyen}, Andrew and {Haiman}, Zolt{\'a}n},
        title = "{Suppressed Accretion onto Massive Black Hole Binaries Surrounded by Thin Disks}",
      journal = {\apj},
         year = 2025,
        month = may,
       volume = {984},
       number = {2},
          eid = {144},
        pages = {144},
          doi = {10.3847/1538-4357/adc727},
archivePrefix = {arXiv},
       eprint = {2410.03830},
 primaryClass = {astro-ph.GA},
       adsurl = {https://ui.adsabs.harvard.edu/abs/2025ApJ...984..144T}
}

@ARTICLE{ONeillTiede:2025,
       author = {{O'Neill}, David and {Tiede}, Christopher and {D'Orazio}, Daniel J. and {Haiman}, Zolt{\'a}n and {MacFadyen}, Andrew},
        title = "{Gravitational Wave Decoupling in Retrograde Circumbinary Disks}",
      journal = {\apj},
         year = 2025,
        month = nov,
       volume = {993},
       number = {2},
          eid = {206},
        pages = {206},
          doi = {10.3847/1538-4357/ae0ca8},
archivePrefix = {arXiv},
       eprint = {2501.11679},
 primaryClass = {astro-ph.HE},
       adsurl = {https://ui.adsabs.harvard.edu/abs/2025ApJ...993..206O}
}

@ARTICLE{Betancourt:2026,
       author = {{Betancourt}, Leonardo and {MacFadyen}, Andrew and {Zrake}, Jonathan},
        title = "{Eccentric Disks from Circumbinary Rings}",
      journal = {arXiv e-prints},
         year = 2026,
        month = jan,
          eid = {arXiv:2601.00741},
        pages = {arXiv:2601.00741},
archivePrefix = {arXiv},
       eprint = {2601.00741},
 primaryClass = {astro-ph.HE},
       adsurl = {https://ui.adsabs.harvard.edu/abs/2026arXiv260100741B}
}

@ARTICLE{LSST:ScienceBook:2009,
       author = {{LSST Science Collaboration} and {Abell}, Paul A. and {Allison}, Julius and {Anderson}, Scott F. and {Andrew}, John R. and {Angel}, J. Roger P. and {Armus}, Lee and {Arnett}, David and {Asztalos}, S.~J. and {Axelrod}, Tim S. and {Bailey}, Stephen and {Ballantyne}, D.~R. and {Bankert}, Justin R. and {Barkhouse}, Wayne A. and {Barr}, Jeffrey D. and {Barrientos}, L. Felipe and {Barth}, Aaron J. and {Bartlett}, James G. and {Becker}, Andrew C. and {Becla}, Jacek and {Beers}, Timothy C. and {Bernstein}, Joseph P. and {Biswas}, Rahul and {Blanton}, Michael R. and {Bloom}, Joshua S. and {Bochanski}, John J. and {Boeshaar}, Pat and {Borne}, Kirk D. and {Bradac}, Marusa and {Brandt}, W.~N. and {Bridge}, Carrie R. and {Brown}, Michael E. and {Brunner}, Robert J. and {Bullock}, James S. and {Burgasser}, Adam J. and {Burge}, James H. and {Burke}, David L. and {Cargile}, Phillip A. and {Chandrasekharan}, Srinivasan and {Chartas}, George and {Chesley}, Steven R. and {Chu}, You-Hua and {Cinabro}, David and {Claire}, Mark W. and {Claver}, Charles F. and {Clowe}, Douglas and {Connolly}, A.~J. and {Cook}, Kem H. and {Cooke}, Jeff and {Cooray}, Asantha and {Covey}, Kevin R. and {Culliton}, Christopher S. and {de Jong}, Roelof and {de Vries}, Willem H. and {Debattista}, Victor P. and {Delgado}, Francisco and {Dell'Antonio}, Ian P. and {Dhital}, Saurav and {Di Stefano}, Rosanne and {Dickinson}, Mark and {Dilday}, Benjamin and {Djorgovski}, S.~G. and {Dobler}, Gregory and {Donalek}, Ciro and {Dubois-Felsmann}, Gregory and {Durech}, Josef and {Eliasdottir}, Ardis and {Eracleous}, Michael and {Eyer}, Laurent and {Falco}, Emilio E. and {Fan}, Xiaohui and {Fassnacht}, Christopher D. and {Ferguson}, Harry C. and {Fernandez}, Yanga R. and {Fields}, Brian D. and {Finkbeiner}, Douglas and {Figueroa}, Eduardo E. and {Fox}, Derek B. and {Francke}, Harold and {Frank}, James S. and {Frieman}, Josh and {Fromenteau}, Sebastien and {Furqan}, Muhammad and {Galaz}, Gaspar and {Gal-Yam}, A. and {Garnavich}, Peter and {Gawiser}, Eric and {Geary}, John and {Gee}, Perry and {Gibson}, Robert R. and {Gilmore}, Kirk and {Grace}, Emily A. and {Green}, Richard F. and {Gressler}, William J. and {Grillmair}, Carl J. and {Habib}, Salman and {Haggerty}, J.~S. and {Hamuy}, Mario and {Harris}, Alan W. and {Hawley}, Suzanne L. and {Heavens}, Alan F. and {Hebb}, Leslie and {Henry}, Todd J. and {Hileman}, Edward and {Hilton}, Eric J. and {Hoadley}, Keri and {Holberg}, J.~B. and {Holman}, Matt J. and {Howell}, Steve B. and {Infante}, Leopoldo and {Ivezic}, Zeljko and {Jacoby}, Suzanne H. and {Jain}, Bhuvnesh and {R} and {Jedicke} and {Jee}, M. James and {Garrett Jernigan}, J. and {Jha}, Saurabh W. and {Johnston}, Kathryn V. and {Jones}, R. Lynne and {Juric}, Mario and {Kaasalainen}, Mikko and {Styliani} and {Kafka} and {Kahn}, Steven M. and {Kaib}, Nathan A. and {Kalirai}, Jason and {Kantor}, Jeff and {Kasliwal}, Mansi M. and {Keeton}, Charles R. and {Kessler}, Richard and {Knezevic}, Zoran and {Kowalski}, Adam and {Krabbendam}, Victor L. and {Krughoff}, K. Simon and {Kulkarni}, Shrinivas and {Kuhlman}, Stephen and {Lacy}, Mark and {Lepine}, Sebastien and {Liang}, Ming and {Lien}, Amy and {Lira}, Paulina and {Long}, Knox S. and {Lorenz}, Suzanne and {Lotz}, Jennifer M. and {Lupton}, R.~H. and {Lutz}, Julie and {Macri}, Lucas M. and {Mahabal}, Ashish A. and {Mandelbaum}, Rachel and {Marshall}, Phil and {May}, Morgan and {McGehee}, Peregrine M. and {Meadows}, Brian T. and {Meert}, Alan and {Milani}, Andrea and {Miller}, Christopher J. and {Miller}, Michelle and {Mills}, David and {Minniti}, Dante and {Monet}, David and {Mukadam}, Anjum S. and {Nakar}, Ehud and {Neill}, Douglas R. and {Newman}, Jeffrey A. and {Nikolaev}, Sergei and {Nordby}, Martin and {O'Connor}, Paul and {Oguri}, Masamune and {Oliver}, John and {Olivier}, Scot S. and {Olsen}, Julia K. and {Olsen}, Knut and {Olszewski}, Edward W. and {Oluseyi}, Hakeem and {Padilla}, Nelson D. and {Parker}, Alex and {Pepper}, Joshua and {Peterson}, John R. and {Petry}, Catherine and {Pinto}, Philip A. and {Pizagno}, James L. and {Popescu}, Bogdan and {Prsa}, Andrej and {Radcka}, Veljko and {Raddick}, M. Jordan and {Rasmussen}, Andrew and {Rau}, Arne and {Rho}, Jeonghee and {Rhoads}, James E. and {Richards}, Gordon T. and {Ridgway}, Stephen T. and {Robertson}, Brant E. and {Roskar}, Rok and {Saha}, Abhijit and {Sarajedini}, Ata and {Scannapieco}, Evan and {Schalk}, Terry and {Schindler}, Rafe and {Schmidt}, Samuel},
        title = "{LSST Science Book, Version 2.0}",
      journal = {arXiv e-prints},
         year = 2009,
        month = dec,
          eid = {arXiv:0912.0201},
        pages = {arXiv:0912.0201},
          doi = {10.48550/arXiv.0912.0201},
archivePrefix = {arXiv},
       eprint = {0912.0201},
 primaryClass = {astro-ph.IM},
       adsurl = {https://ui.adsabs.harvard.edu/abs/2009arXiv0912.0201L}
}

@ARTICLE{Kelley:LensingDetect:2021,
       author = {{Kelley}, Luke Zoltan and {D'Orazio}, Daniel J. and {Di Stefano}, Rosanne},
        title = "{Gravitational self-lensing in populations of massive black hole binaries}",
      journal = {\mnras},
         year = 2021,
        month = dec,
       volume = {508},
       number = {2},
        pages = {2524-2536},
          doi = {10.1093/mnras/stab2776},
archivePrefix = {arXiv},
       eprint = {2107.07522},
 primaryClass = {astro-ph.HE},
       adsurl = {https://ui.adsabs.harvard.edu/abs/2021MNRAS.508.2524K}
}

@ARTICLE{Berlok19,
       author = {{Berlok}, Thomas and {Pfrommer}, Christoph},
        title = "{On the Kelvin-Helmholtz instability with smooth initial conditions - linear theory and simulations}",
      journal = {\mnras},
         year = 2019,
        month = may,
       volume = {485},
       number = {1},
        pages = {908-923},
          doi = {10.1093/mnras/stz379},
archivePrefix = {arXiv},
       eprint = {1902.01403},
 primaryClass = {astro-ph.GA},
       adsurl = {https://ui.adsabs.harvard.edu/abs/2019MNRAS.485..908B}
}

@ARTICLE{Tiwari:2025,
       author = {{Tiwari}, Vishal and {Chan}, Chi-Ho and {Bogdanovi{\'c}}, Tamara and {Jiang}, Yan-Fei and {Davis}, Shane W. and {Ferrel}, Simon},
        title = "{Radiation Magnetohydrodynamic Simulation of Sub-Eddington Circumbinary Disk around an Equal-mass Massive Black Hole Binary}",
      journal = {\apj},
         year = 2025,
        month = jun,
       volume = {986},
       number = {2},
          eid = {158},
        pages = {158},
          doi = {10.3847/1538-4357/add408},
archivePrefix = {arXiv},
       eprint = {2502.18584},
 primaryClass = {astro-ph.HE},
       adsurl = {https://ui.adsabs.harvard.edu/abs/2025ApJ...986..158T}
}

@ARTICLE{NarayanYi:ADAF:1994,
       author = {{Narayan}, Ramesh and {Yi}, Insu},
        title = "{Advection-dominated Accretion: A Self-similar Solution}",
      journal = {\apjl},
         year = 1994,
        month = jun,
       volume = {428},
        pages = {L13},
          doi = {10.1086/187381},
archivePrefix = {arXiv},
       eprint = {astro-ph/9403052},
 primaryClass = {astro-ph},
       adsurl = {https://ui.adsabs.harvard.edu/abs/1994ApJ...428L..13N}
}

@ARTICLE{Tiede:badaf:2025,
       author = {{Tiede}, Christopher and {D'Orazio}, Daniel J.},
        title = "{Hot, Cold, and Multicomponent Accretion Flows around Supermassive Black Hole Binaries}",
      journal = {\apj},
         year = 2025,
        month = dec,
       volume = {995},
       number = {1},
          eid = {68},
        pages = {68},
          doi = {10.3847/1538-4357/ae17ba},
archivePrefix = {arXiv},
       eprint = {2508.11748},
 primaryClass = {astro-ph.HE},
       adsurl = {https://ui.adsabs.harvard.edu/abs/2025ApJ...995...68T}
}

@ARTICLE{LyndenBell-Pringle:1974,
       author = {{Lynden-Bell}, D. and {Pringle}, J.~E.},
        title = "{The evolution of viscous discs and the origin of the nebular variables.}",
      journal = {\mnras},
         year = 1974,
        month = sep,
       volume = {168},
        pages = {603-637},
          doi = {10.1093/mnras/168.3.603},
       adsurl = {https://ui.adsabs.harvard.edu/abs/1974MNRAS.168..603L}
}

@ARTICLE{Roman2015,
       author = {{Spergel}, D. and {Gehrels}, N. and {Baltay}, C. and {Bennett}, D. and {Breckinridge}, J. and {Donahue}, M. and {Dressler}, A. and {Gaudi}, B.~S. and {Greene}, T. and {Guyon}, O. and {Hirata}, C. and {Kalirai}, J. and {Kasdin}, N.~J. and {Macintosh}, B. and {Moos}, W. and {Perlmutter}, S. and {Postman}, M. and {Rauscher}, B. and {Rhodes}, J. and {Wang}, Y. and {Weinberg}, D. and {Benford}, D. and {Hudson}, M. and {Jeong}, W.-S. and {Mellier}, Y. and {Traub}, W. and {Yamada}, T. and {Capak}, P. and {Colbert}, J. and {Masters}, D. and {Penny}, M. and {Savransky}, D. and {Stern}, D. and {Zimmerman}, N. and {Barry}, R. and {Bartusek}, L. and {Carpenter}, K. and {Cheng}, E. and {Content}, D. and {Dekens}, F. and {Demers}, R. and {Grady}, K. and {Jackson}, C. and {Kuan}, G. and {Kruk}, J. and {Melton}, M. and {Nemati}, B. and {Parvin}, B. and {Poberezhskiy}, I. and {Peddie}, C. and {Ruffa}, J. and {Wallace}, J.~K. and {Whipple}, A. and {Wollack}, E. and {Zhao}, F.},
        title = "{Wide-Field InfrarRed Survey Telescope-Astrophysics Focused Telescope Assets WFIRST-AFTA 2015 Report}",
      journal = {arXiv e-prints},
         year = 2015,
        month = mar,
          eid = {arXiv:1503.03757},
        pages = {arXiv:1503.03757},
          doi = {10.48550/arXiv.1503.03757},
archivePrefix = {arXiv},
       eprint = {1503.03757},
 primaryClass = {astro-ph.IM},
       adsurl = {https://ui.adsabs.harvard.edu/abs/2015arXiv150303757S}
}

@ARTICLE{Grcic:2026,
       author = {{Grci{\'c}}, Marcela and {D'Orazio}, Daniel J. and {Pessah}, Martin E.},
        title = "{Insights from Analytical Theory of Eccentric Circumbinary Disks}",
      journal = {\apj},
         year = 2026,
        month = feb,
       volume = {998},
       number = {1},
          eid = {4},
        pages = {4},
          doi = {10.3847/1538-4357/ae29b0},
archivePrefix = {arXiv},
       eprint = {2504.17658},
 primaryClass = {astro-ph.SR},
       adsurl = {https://ui.adsabs.harvard.edu/abs/2026ApJ...998....4G}
}

@ARTICLE{DOrazioCharisi:Review:2023,
       author = {{D'Orazio}, Daniel J. and {Charisi}, Maria},
        title = "{Observational Signatures of Supermassive Black Hole Binaries}",
      journal = {arXiv e-prints},
         year = 2023,
        month = oct,
          eid = {arXiv:2310.16896},
        pages = {arXiv:2310.16896},
          doi = {10.48550/arXiv.2310.16896},
archivePrefix = {arXiv},
       eprint = {2310.16896},
 primaryClass = {astro-ph.HE},
       adsurl = {https://ui.adsabs.harvard.edu/abs/2023arXiv231016896D}
}

@ARTICLE{Fan:FirstWLQ:1999,
       author = {{Fan}, Xiaohui and {Strauss}, Michael A. and {Gunn}, James E. and {Lupton}, Robert H. and {Carilli}, C.~L. and {Rupen}, M.~P. and {Schmidt}, Gary D. and {Moustakas}, Leonidas A. and {Davis}, Marc and {Annis}, James and et al.},
        title = "{The Discovery of a High-Redshift Quasar without Emission Lines from Sloan Digital Sky Survey Commissioning Data}",
      journal = {\apjl},
         year = 1999,
        month = dec,
       volume = {526},
       number = {2},
        pages = {L57-L60},
          doi = {10.1086/312382},
archivePrefix = {arXiv},
       eprint = {astro-ph/9910001},
 primaryClass = {astro-ph},
       adsurl = {https://ui.adsabs.harvard.edu/abs/1999ApJ...526L..57F}
}

@ARTICLE{Dioamond-Stanic:WLQs:2009,
       author = {{Diamond-Stanic}, Aleksandar M. and {Fan}, Xiaohui and {Brandt}, W.~N. and {Shemmer}, Ohad and {Strauss}, Michael A. and {Anderson}, Scott F. and {Carilli}, Christopher L. and {Gibson}, Robert R. and {Jiang}, Linhua and {Kim}, J. Serena and et al.},
        title = "{High-redshift SDSS Quasars with Weak Emission Lines}",
      journal = {\apj},
         year = 2009,
        month = jul,
       volume = {699},
       number = {1},
        pages = {782-799},
          doi = {10.1088/0004-637X/699/1/782},
archivePrefix = {arXiv},
       eprint = {0904.2181},
 primaryClass = {astro-ph.GA},
       adsurl = {https://ui.adsabs.harvard.edu/abs/2009ApJ...699..782D}
}

@ARTICLE{Plotkin:SDSS-WLQs:2010,
       author = {{Plotkin}, Richard M. and {Anderson}, Scott F. and {Brandt}, W.~N. and {Diamond-Stanic}, Aleksandar M. and {Fan}, Xiaohui and {Hall}, Patrick B. and {Kimball}, Amy E. and {Richmond}, Michael W. and {Schneider}, Donald P. and {Shemmer}, Ohad and et al.},
        title = "{Optically Selected BL Lacertae Candidates from the Sloan Digital Sky Survey Data Release Seven}",
      journal = {\aj},
         year = 2010,
        month = feb,
       volume = {139},
       number = {2},
        pages = {390-414},
          doi = {10.1088/0004-6256/139/2/390},
archivePrefix = {arXiv},
       eprint = {0911.0423},
 primaryClass = {astro-ph.CO},
       adsurl = {https://ui.adsabs.harvard.edu/abs/2010AJ....139..390P}
}

@ARTICLE{Wu:XrayWeakQuasars:2011,
       author = {{Wu}, Jianfeng and {Brandt}, W.~N. and {Hall}, Patrick B. and {Gibson}, Robert R. and {Richards}, Gordon T. and {Schneider}, Donald P. and {Shemmer}, Ohad and {Just}, Dennis W. and {Schmidt}, Sarah J.},
        title = "{A Population of X-Ray Weak Quasars: PHL 1811 Analogs at High Redshift}",
      journal = {\apj},
         year = 2011,
        month = jul,
       volume = {736},
       number = {1},
          eid = {28},
        pages = {28},
          doi = {10.1088/0004-637X/736/1/28},
archivePrefix = {arXiv},
       eprint = {1104.3861},
 primaryClass = {astro-ph.CO},
       adsurl = {https://ui.adsabs.harvard.edu/abs/2011ApJ...736...28W}
}

@ARTICLE{Meusinger:WLQs:2014,
       author = {{Meusinger}, H. and {Balafkan}, N.},
        title = "{A large sample of Kohonen-selected SDSS quasars with weak emission lines: selection effects and statistical properties}",
      journal = {\aap},
         year = 2014,
        month = aug,
       volume = {568},
          eid = {A114},
        pages = {A114},
          doi = {10.1051/0004-6361/201423810},
archivePrefix = {arXiv},
       eprint = {1407.0193},
 primaryClass = {astro-ph.GA},
       adsurl = {https://ui.adsabs.harvard.edu/abs/2014A&A...568A.114M}
}

@article{numpy,
  author = {{Harris}, Charles R. and others},
  title = {Array programming with NumPy},
  journal = {Nature},
  volume = {585},
  pages = {357--362},
  year = {2020},
  doi = {10.1038/s41586-020-2649-2}
}

@article{scipy,
  author = {{Virtanen}, Pauli and others},
  title = {SciPy 1.0: Fundamental Algorithms for Scientific Computing in Python},
  journal = {Nature Methods},
  volume = {17},
  pages = {261--272},
  year = {2020},
  doi = {10.1038/s41592-019-0686-2}
}

@article{matplotlib,
  author = {{Hunter}, J. D.},
  title = {Matplotlib: A 2D Graphics Environment},
  journal = {Computing in Science and Engineering},
  volume = {9},
  number = {3},
  pages = {90--95},
  year = {2007},
  doi = {10.1109/MCSE.2007.55}
}

@article{cmasher,
  author = {{van der Velden}, Ellert},
  title = {CMasher: Scientific colormaps for making accessible, informative and 'cmashing' plots},
  journal = {Journal of Open Source Software},
  volume = {5},
  number = {46},
  pages = {2004},
  year = {2020},
  doi = {10.21105/joss.02004}
}

@ARTICLE{Gammie:2001,
       author = {{Gammie}, Charles F.},
        title = "{Nonlinear Outcome of Gravitational Instability in Cooling, Gaseous Disks}",
      journal = {\apj},
         year = 2001,
        month = may,
       volume = {553},
       number = {1},
        pages = {174-183},
          doi = {10.1086/320631},
archivePrefix = {arXiv},
       eprint = {astro-ph/0101501},
 primaryClass = {astro-ph},
       adsurl = {https://ui.adsabs.harvard.edu/abs/2001ApJ...553..174G}
}
\end{document}